%% file: thesis.tex
\documentclass[]{macro/neu_msthesis}

\title{SMASH: Sparse Matrix Atomic Scratchpad Hashing}

\author{Kaustubh Shivdikar}

\dept{Electrical and Computer Engineering}

\degreename{Electrical and Computer Engineering}

\field{Electrical and Computer Engineering}

\submitdate{April 2021}

\numberofmembers{4}

\principaladviser{Dr. David Kaeli}
\firstreader{Dr. Devesh Tiwari}
\secondreader{Dr. Fabrizio Petrini}

\thirdreader{Dr. Fabio Checconi}
\fourthreader{Fourth Reader Name}
\fifthreader{Fifth Reader Name}

\chairman{Dr. Srinivas Tadigadapa}

\dean{Dr. Waleed Meleis}

\input{macro/macro}

\begin{document}

\pdfbookmark[1]{Cover}{cover}

\titlepage

\begin{frontmatter}
\input{tex/dedication.tex}
\pdfbookmark[1]{Table of Contents}{contents}
\tableofcontents
\listoffigures
\newpage\ssp
\listoftables

\input{tex/acronyms.tex}

\input{tex/acknowledgements.tex}

\input{tex/abstract.tex}

\end{frontmatter}


\pagestyle{headings}

\input{tex/intro.tex}

\input{tex/background.tex}

\input{tex/related_work.tex}

\input{tex/architecture.tex}

\input{tex/smash.tex}

\input{tex/results.tex}


\input{tex/conclusion.tex}

\bibliographystyle{plain}

\bibliography{bib/thesis}

\appendix

\printindex

\end{document}

%% file: macro/macro.tex
\usepackage{amsfonts}			
\usepackage{times}		

\usepackage{algorithmic}
\usepackage[linesnumbered,algoruled,boxed,lined]{algorithm2e}
\usepackage[dvipsnames]{xcolor}
\usepackage{setspace}
\usepackage{pbox}
\usepackage[super]{nth}
\usepackage{placeins}

\newcommand{\barrier}[0]{\textcolor{red}{\textbf{barrier}}}

\newcommand{\ifno}[1]{}

\usepackage{multirow}
\makeindex
\usepackage{makeidx}
\usepackage{acronym}
\usepackage{amsmath}

\usepackage{url}
\ifnum\pdfoutput>0
\usepackage[pdftex]{graphicx}
\else
\usepackage{graphicx}
\fi

\ifnum\pdfoutput>0
\usepackage[pdftex,
bookmarks=true,
bookmarksnumbered=true,
hypertexnames=false,
breaklinks=true           
]{hyperref}
\else
\usepackage[hypertex,
bookmarks=true,
bookmarksnumbered=true,
hypertexnames=false,
breaklinks=true           
]{hyperref}
\fi

\hypersetup{				
pdfauthor = {\authorRef},
pdftitle = {\titleRef},
pdfsubject = {\expandafter{\degreeRef} thesis submitted to Northeastern University},
pdfkeywords = {add keywords here}
pdfcreator = {LaTeX with hyperref package},
pdfproducer = {dvips + ps2pdf}}

%% file: tex/dedication.tex

\begin{dedication}

To the people trying to make the architectures of today, history tomorrow.

\end{dedication}

%% file: tex/acronyms.tex

\chapter*{List of Acronyms}
\addcontentsline{toc}{chapter}{List of Acronyms}

\begin{acronym}

\acro{ABM}{\textit{Advanced Bit Manipulation}}

\acro{ATT}{Address translation tables}

\acro{ADX}{\textit{Multiprecision Add Carry}}

\acro{AESI}{\textit{Advanced Encryption Instructions}}

\acro{AMD}{\textit{Advanced Micro Devices}}

\acro{AMD}{\textit{Central Processing Unit}}

\acro{DGAS}{\textit{Distributed Global Address Space}}

\acro{AVX2}{\textit{Advanced Vector Instructions 2}}

\acro{BFS}{\textit{Breadth First Search}}

\acro{BLAS}{\textit{Basic Linear Algebra Subprograms}}

\acro{BMI2}{\textit{Bit Manipulation Instruction Set 2}}

\acro{C-RAM}{\textit{Computational RAM}}

\acro{ELL}{\textit{ELLPACK}}

\acro{EMMX}{\textit{Extended MMX Extension}}

\acro{CLMUL}{\textit{Carry-less Multiplication Extension}}

\acro{CPU}{\textit{Central Processing Unit}}

\acro{CSR}{\textit{Compressed Sparse Row}}

\acro{CSC}{\textit{Compressed Sparse Column}}

\acro{DARPA}{Defense Advanced Research Projects Agency}

\acro{DFFT}{\textit{Dense Fast Fourier Transform}}

\acro{DMA}{\textit{Direct Memory Access}}

\acro{DP}{\textit{Double Precision}}

\acro{DRAM}{\textit{Dynamic Random Access Memory}}

\acro{F16C}{\textit{16-bit Floating Point Conversion}}

\acro{FMA}{\textit{Floating-point Multiply and Add}}

\acro{FMA3}{\textit{3-Operand Fused-Multiply-Add}}

\acro{GCN}{\textit{Graph Convolutional Network}}

\acro{GEMM}{\textit{General Matrix Multiplication}}

\acro{GNN}{\textit{Graph Neural Networks}}

\acro{GPGPU}{\textit{General-Purpose Graphic Processing Unit}}

\acro{GPU}{\textit{Graphics Processing Unit}}

\acro{HBM}{\textit{High Bandwidth Memory}}

\acro{HP}{\textit{Half Precision}}

\acro{HIVE}{Hierarchical Identify Verify Exploit}

\acro{IPC}{\textit{Instructions per Cycle}}

\acro{ISA}{\textit{Instruction Set Architecture}}

\acro{MLP}{\textit{Multi-layer Perceptron}}

\acro{MKL}{\textit{Math Kernel Library}}

\acro{MLP}{\textit{Multi-Layer Perceptrons}}

\acro{MTC}{\textit{Multi-threaded Core}}

\acro{PIUMA}{Programmable Integrated Unified Memory Architecture}

\acro{RF}{Register File}

\acro{SP}{\textit{Single Precision}}

\acro{OMP}{\textit{OpenMP}}

\acro{OS}{\textit{Operating System}}

\acro{PIM}{\textit{Processor in Memory}}

\acro{QP}{\textit{Quadruple Precision}}

\acro{RMAT}{\textit{Recursive Matrix}}

\acro{SFFT}{\textit{Sparse Fast Fourier Transform}}

\acro{SIMD}{\textit{Single Instruction Multiple Data}}

\acro{SMASH}{\textit{Sparse Matrix Atomic Scratchpad Hashing}}

\acro{SPAD}{\textit{Scratchpad}}

\acro{SPMD}{\textit{Single Program, Multiple Data}}

\acro{SpGEMM}{\textit{Sparse Matrix-Matrix Multiply}}

\acro{SpMV}{\textit{Sparse Matrix-Vector}}

\acro{SSE4.2}{\textit{Streaming SIMD Extensions 4.2}}

\acro{STC}{\textit{Single-threaded Core}}

\end{acronym}

%% file: tex/acknowledgements.tex

\begin{acknowledgements}

I would like to start with thanking my Ph.D. advisor, Prof. David Kaeli, for his dedicated support and motivation in this thesis as well as the project.
Would like to thank Dr. Fabrizio Petrini for providing me with this extra-ordinary opportunity to experiment with the latest and greatest tools as well as his guidance on SpGEMM.
In addition, would like to thank Dr. Fabio Checconi for his valuable feedback and inspiration for SMASH kernels.
Thank you Intel for exposing me to some of the brightest minds in the valley.
Thank you all members of NUCAR, Dr. Norm Rubin, Dr. Yifan Sun, Elmira Karimi, Malith Jayaweera, Zlatan Feric, Derek Rodriguez, Julian Gutierrez, Trinayan Baruah, and Yuhui Bao for your guidance and support.
A special thanks to thank Nicolas Agostini, the source of inspiration and motivation, throughout this project.

With the world coming to a grinding halt with an epidemic that showed no signs of an end, I would like to thank my parents Chandrakant and Anagha and my brother Saumil for the relentless support in all aspects over these last years.
Last but not least, I would like to thank all my friends in US for their constant motivation, who succeeded in making this journey memorable.

\end{acknowledgements}

%% file: tex/abstract.tex

\begin{abstract}

\paragraph{Abstract:} Sparse matrices, more specifically \ac{SpGEMM} kernels, are commonly found in a wide range of applications, spanning graph-based path-finding to machine learning algorithms (e.g., neural networks). A particular challenge in implementing \ac{SpGEMM} kernels has been the pressure placed on DRAM memory. One approach to tackle this problem is to use an inner product method for the \ac{SpGEMM} kernel implementation. While the inner product produces fewer intermediate results, it can end up saturating the memory bandwidth, given the high number of redundant fetches of the input matrix elements. Using an outer product-based \ac{SpGEMM} kernel can reduce redundant fetches, but at the cost of increased overhead due to extra computation and memory accesses for producing/managing partial products.

In this thesis, we introduce a novel \ac{SpGEMM} kernel implementation based on the row-wise product approach. We leverage atomic instructions to merge intermediate partial products as they are generated. The use of atomic instructions eliminates the need to create partial product matrices, thus eliminating redundant DRAM fetches.

To evaluate our row-wise product approach, we map an optimized \ac{SpGEMM} kernel to a custom accelerator designed to accelerate graph-based applications. The targeted accelerator is an experimental system named PIUMA, being developed by Intel. PIUMA provides several attractive features, including fast context switching, user-configurable caches, globally addressable memory, non-coherent caches, and asynchronous pipelines.  We tailor our \ac{SpGEMM} kernel to exploit many of the features of the PIUMA fabric.

This thesis compares our \ac{SpGEMM} implementation against prior solutions, all mapped to the PIUMA framework. We briefly describe some of the PIUMA architecture features and then delve into the details of our optimized \ac{SpGEMM} kernel.  Our \ac{SpGEMM} kernel can achieve $9.4\times$ speedup as compared to competing approaches.
\end{abstract}

%% file: tex/intro.tex

\chapter{Introduction}
\label{chap:intro}

\section{The Development of Matrix-based Methods}
\label{chap:intro:history}

In 1812, a French mathematician named Jacques Philippe Marie Binet pointed out several important computations involved the multiplication of two matrices~\cite{knill_2009}. On November 30 of the same year, he provided a lecture on his observation and further extended his work, leading to the Cauchy-Binet formula~\cite{knill2014cauchy}. This is one of the oldest known sources of the discovery of matrix multiplication. Matrix multiplication was described as a method of multiply data arranged in rows.
Later, in the year 1850, Arthur Cayley applied matrix multiplication to solve a system of linear equations~\cite{britannica}, showing applications of this idea to solve an important class of mathematical problems.

\section{High Performance Computing and Matrices}
\label{chap:intro:standardization}

The \nth{20} century witnessed developments in computer technology. Computers, which were initially developed to crunch numbers for tabulating the United States census~\cite{zimmermann_2017}, were soon being used to perform
calculations for a variety of physics and mathematics problems, many that included matrix multiplication~\cite{zimmermann_2017}.

Use-cases for matrix multiplication applications were so widespread that they demanded standardization \cite{kaagstrom1998gemm}. In 1979, the BLAS Technical forum published a report on standardization of a few of the common linear algebra operations (also known as subroutines), which they referred to as \textit{Level-1 Basic Linear Algebra Subroutines} (BLAS) \cite{lawson1979basic}. Later, in 1986 and 1988, BLAS was further augmented with Level-2 and Level-3 subroutines, respectively. The Level-3 subroutines included the matrix multiplication subroutine (also known as GEMM kernel). The \ac{GEMM} kernel implementations were designed to work with dense matrices (matrices with mostly non-zero elements).

Matrix multiplication is a key operation in many scientific computations. One such computation is graph analysis. Graph analysis commonly represents graphs using an adjacency matrix and then performs matrix multiplication operations on these matrices. The associated adjacency matrices are inherently sparse \cite{delorimier2006graphstep} (with very few non-zeros) due to many graphs' structures.  Early library implementations of \ac{GEMM} performed poorly on such sparse matrices. By 2002, the BLAS Technical forum adopted a new standard for such sparse data~\cite{duff2002overview}. They presented the \textit{Sparse Basic Linear Algebra Subprograms} (Sparse BLAS), which included subroutines that included \ac{SpGEMM} (also known as the \ac{SpGEMM} kernel), and focused on optimizations required for sparse matrix multiplication \cite{duff2002overview}.

\section{Applications}
\label{chap:intro:applications}

Today, \ac{GEMM} and \ac{SpGEMM} kernels have found their way into many important applications. Some of these applications include:

\begin{enumerate}
	\item Data encryption: AES, SHA1, SHA2, Twofish \cite{canright2008very, ivanov2016autonomous, daemen1999rijndael, murphy2002essential, daemen1999aes, trichina2002simplified, li2005efficient, thakkar2017video};
	\item Data compression and Decompression: zip files, JPEG and PNG image compression~\cite{pennebaker1992jpeg, lin2003efficient};
	\item Image processing: filters for real-time image processing, such as Sobel filtering, image sharpening, image blurring~\cite{pradabpet2009efficient, ndour2018evaluation, shah2013performance, shivdikar2015automatic};
	\item Pathfinding: \ac{BFS} and Dijkstra's Algorithm \cite{zwick2002all};
	\item Signal processing: \ac{DFFT}, \ac{SFFT} \cite{hsieh2013sparse, schumacher2013high, pitsianis2005efficient};
	\item Simulations: N-body, raytracing, and Monte-Carlo~\cite{aarseth2003gravitational, stock2008toward, yokota2012hierarchical, wilkinson2004sparse}; and
	\item Machine learning: various supervised and unsupervised learning algorithms are implemented using GEMM kernels~Deep learning utilizes GEMM kernels for convolution layers \cite{park2018deep, moss2018customizable, chetlur2014cudnn, osawa2017accelerating, tschannen2018strassennets, alzantot2017rstensorflow, shazeer2018mesh, vasudevan2017parallel, wurfl2016deep}.
\end{enumerate}

This thesis's motivation lies in improving the performance of \ac{SpGEMM} kernels, which will have a significant impact on many important applications.

Recently, we have seen growth in graph-based applications in the industry.

Facebook \cite{bailey2018social}, Google~\cite{48449}, Twitter~\cite{info:doi/10.2196/jmir.2741}, Amazon~\cite{smith2017two}, Netflix~\cite{bennett2007netflix}, Cora~\cite{kipf2016variational}, Citeseer~\cite{giles1998citeseer} and many other large companies use graphs to analyze social networks, citation networks, and even recommend products.

The currently available computational frameworks have not kept up with the ever-increasing demands of graph-based workloads~\cite{wang2020gnnadvisor}.  One of the key components of such applications is the sparse matrix multiplication kernel (SpGEMM kernel). As compared to their dense counterparts, SpGEMM kernels are complex and harder to optimize. Traditional multi-core CPUs and many-core GPU architectures provide limited performance over SpGEMM kernels, mainly due to their irregular memory access patterns and unbalanced work distribution.

\section{Motivation}
\label{chap:intro:motivation}

One graph-analysis application that is growing in popularity is the \ac{GNN}. GNNs represent features of a node in the graph with a vector. These vectors are then recursively aggregated and transformed, based on the features of the neighboring nodes~\cite{xu2018powerful}. These features can then be used to classify nodes or perform inference on datasets. Unlike traditional neural networks that work with dense data structures, such as \ac{MLP}, \ac{GNN}s operate on sparse graph structures~\cite{baruah2021gnnmark}. They have been becoming increasingly popular, given their high accuracy for node classification on graph-structured datasets~\cite{zhang2020architectural, zhou2018graph}.
\begin{figure}[htbp]
	\centering
	\includegraphics[width=0.9\textwidth]{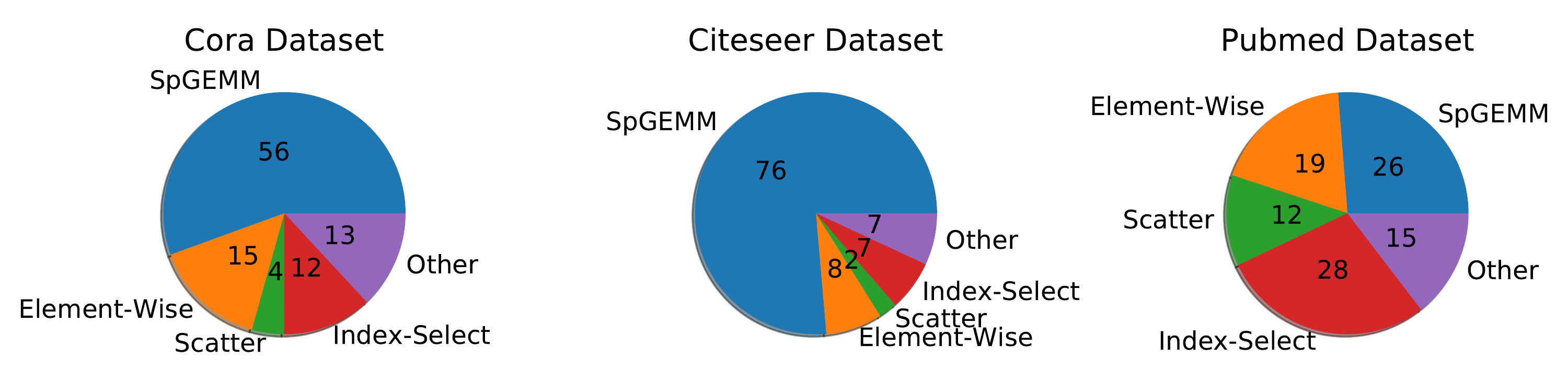}
	\caption[Deep learning Workload characterization]{\label{fig:spgemm_characterization} Graph Convolutional Network (GCN) kernel execution time breakdown.}
\end{figure}
From a computational perspective, \ac{GNN}s are comprised of a mix of kernels, including element-wise operations, transpose operations, dense matrix multiplication, sparse matrix multiplication, index selection, reduction, and batch normalization.
One example of a \ac{GNN} is a Graph Convolutional Network or \ac{GCN}. A GCN is similar to a Convolutional Neural Network (CNN), except that it performs convolution operations on a graph instead of image-based operations on pixels~\cite{chen2020fusegnn}.
Figure~\ref{fig:spgemm_characterization} shows a breakdown of time spent in each of these operations in a \ac{GCN} application.

The time spent in computing \ac{SpGEMM} kernels is a function of two factors: 1) the sparsity of the dataset and 2) the sparsity pattern.  Although it is difficult to compare sparsity patterns, Table~\ref{table:graph_dataset} presents the degree of sparsity for various graph datasets. 
Many workloads, including graph convolution~\cite{chiang2019cluster}, node classification \cite{abu2020n}, path planning~\cite{lugowski2012flexible}, use the datasets shown in Table~\ref{table:graph_dataset} to analyze graphs and derive useful information.
\ac{SpGEMM} remains an integral kernel used in such workloads, processing highly sparse datasets, thus providing us with an opportunity to exploit this sparsity using an optimized kernel.

\begin{table}[ht]
	\centering 
	\begin{tabular}{l r r c}
		\hline\hline 
		\textbf{Dataset} & \textbf{Vertices} & \textbf{Edges} & \textbf{Degree of Sparsity} \\ [0.5ex]
		\hline 
		Citeseer & 3327 & 9464 & 99.914 \\
		Cora & 2708 & 10,858 & 99.851\\
		Pubmed & 19,717 & 88,676 & 99.977 \\ 
		Wikipedia RfA & 113,80 & 188,077 & 99.854 \\ 
		Epinions & 75,888 & 508,837 & 99.991 \\ 
		Slashdot & 82,144 & 549,202 & 99.991 \\
	    Astro Physics Collaborations & 18,772 & 792,320 & 99.775 \\ 
		NotreDame & 325,729 & 1,497,134 & 99.998 \\
		Amazon & 334,863 & 1,851,744 & 99.998 \\
		Google Page Hyperlinks & 916,428 & 5,105,039 & 99.999 \\
		Youtube & 1,134,890 & 11,950,496 & 99.999 \\
		Patent Citations & 3,774,768 & 16,518,948 & 99.999 \\ 
		Stack Overflow & 2,601,977 & 36,233,450 & 99.999 \\
		Orkut & 3,072,441 & 486,740,332 & 99.994 \\
		Twitter Follower Network & 41,652,230 & 1,468,365,182 & 99.999 \\ [1ex] 
		\hline 
	\end{tabular}
	\caption{Sparse Graph datasets} 
	\label{table:graph_dataset} 
\end{table}

Given the dominance of \ac{SpGEMM} execution in  GNN applications, we focus on accelerating \ac{SpGEMM} kernels to reduce their execution time. In this thesis, we focus on identifying the underlying bottlenecks of \ac{SpGEMM} kernels and improving these workloads' performance with datasets with varying degrees of sparsity.  We target the Intel PIUMA parallel accelerator, providing us with a state-of-the-art target to demonstrate our approach's utility.

\section{Dataflow in SpGEMM}
\label{chap:intro:problem}

\begin{figure}[htbp]
	\centering
	\includegraphics[width=1.0\textwidth]{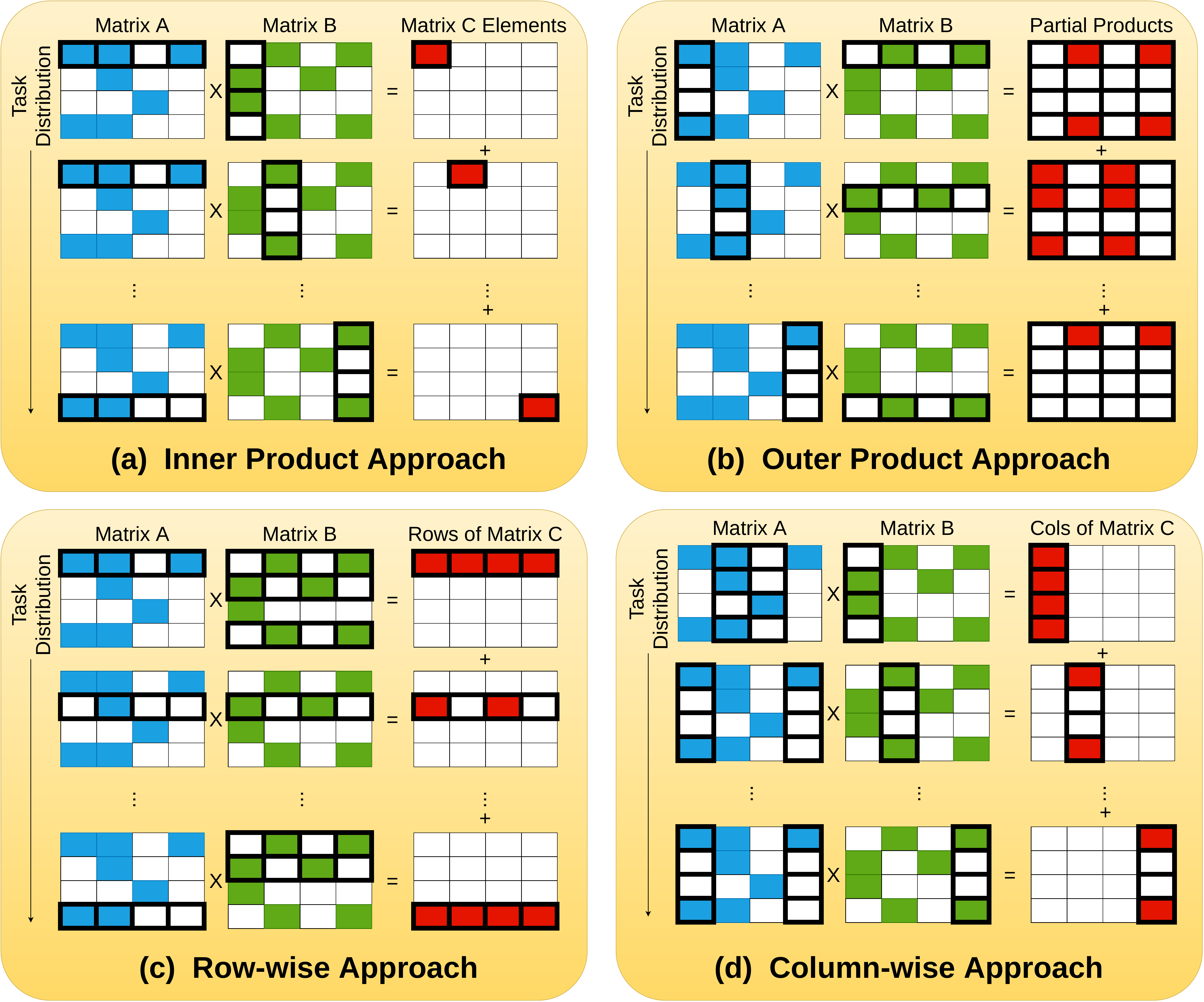}
	\caption[Matrix Multiplication Approaches]{\label{fig:matrix_multiplication} Matrix multiplication approaches}
\end{figure}

There are four common approaches used to multiply two matrices (as shown in Figure~\ref{fig:matrix_multiplication})~\cite{srivastava2020matraptor}:
\begin{enumerate}
    \item Inner product approach: $Row(A) \times Col(B) = Element(C)$
    \item Outer product approach: $Col(A) \times Row(B) = Partial\ Products\ of\ Matrix(C)$
    \item Row-wise product approach: $Row(A) \times Corresponding\ Rows(B) = Row(C)$
    \item Column-wise product approach: $Corresponding\ Cols(A) \times Col(B) = Col(C)$
\end{enumerate}

The most widely used approach for matrix multiplication is the inner product approach.
Inner-product methods are based on computing a dot product of a row of Matrix $A$ with a column of Matrix $B$, generating a single output element of Matrix $C$ (Figure \ref{fig:matrix_multiplication} (a)).

This leads to multiple reads of the input matrices but results in a single write of the output matrix elements.
Thus, inner-product based methods exhibit poor input reuse, but good output reuse. Equation~\ref{eq:inner_product} denotes the operations of the inner product method:

\begin{equation}
c_{i,j} = \sum_{k=0}^{N-1} a_{i,k} \times b_{k,j} 
\label{eq:inner_product}
\end{equation}
where $A$ and $B$ are input matrices, $c_{i,j}$ is an element of the $i^{th}$ row and $j^{th}$ column of the output matrix $C$, $N$ is the number of columns in the $A$ matrix, subsequently $a_{i,k}$ and $b_{k,j}$ are elements of the corresponding rows and columns of matrices $A$ and $B$, respectively.

\begin{table}[ht]
	\centering 
	\begin{tabular}{l c c c c}
		\hline\hline 
		\textbf{Method} & \textbf{Input Reuse} & \textbf{Output Reuse} & \textbf{Intermediate Size} & \textbf{Disadvantage} \\ [0.5ex]
		\hline 
		Inner Product & Poor & Good & Small & Redundant input reads\\
		Outer Product & Good & Poor & Large & Large intermediate size\\
		Row-wise & Poor & Good & Small & Load imbalance\\
		Col-wise & Poor & Good & Small & Load imbalance\\ [1ex]
		\hline 
	\end{tabular}
	\caption{Matrix Multiplication Methods} 
	\label{table:multiplication_methods} 
\end{table}

In contrast, the outer-product method multiplies a single row of Matrix $A$ with all the rows of matrix $B$ to produce partial products (Figure \ref{fig:matrix_multiplication} (b)). These partial products are stored in intermediate matrices and are later merged to form the output matrix~\cite{liu2020systolic, richter2018symbolic}.
This leads to a single read of the input matrices but multiple writes of the partial product output matrices.
Thus, in contrast to the inner-product method, the outer-product method exhibits good input reuse but poor output reuse.
Computation of the output matrix using an outer product approach is expressed in Equation \ref{eq:outer_product}:

\begin{equation}
\begin{aligned}
Output\ Matrix = \sum_{n=0}^{N-1}C_{n}\\
C_{n} = a_{n}b_{n}
\end{aligned}
\label{eq:outer_product}
\end{equation}
where $C_{n}$ is a partial product matrix of output matrix $C$, $A$ and $B$ represent input matrices, $N$ is the number of columns in matrix $A$ and $a_{n}b_{n}$ is a cross-product of $n^{th}$ column of $A$ and $n^{th}$ row of $B$.

The row-wise approach consists of a scalar product of every element of a row of matrix $A$ with corresponding rows of the matrix $B$ (see Figure~\ref{fig:matrix_multiplication} (c) and Equation~\ref{eq:row_wise}).
The case for the column-wise approach is similar, where a single column of the matrix $B$ is multiplied by corresponding columns of matrix $A$ (as seen in Figure~\ref{fig:matrix_multiplication} (d) and Equation~\ref{eq:col_wise}).
Both row-wise and column-wise products have similar dataflow properties. They both exhibit poor input reuse due to redundant accesses made to one of the two input matrices. As opposed to an outer product approach, they do not generate a large number of intermediate products because partial products are immediately merged after generation. Thus, both of these approaches produce high output reuse.
In addition, the inner and outer product approaches require the input matrices to be stored in opposite storage formats, matrix $A$ in row-major and matrix $B$ in column-major for the inner product, and vice versa for the outer product. On the other hand, both row-wise and column-wise require both input matrices to be stored in the same storage format.
A significant disadvantage of the row-wise and column-wise approach is a skewed matrix (matrix with unevenly distributed non-zeros) can cause load imbalance in computations. This problem of load imbalance, and a solution for the same, are further described in Section~\ref{chap:smash:section:tokenization} of this thesis.

\begin{equation}
\begin{aligned}
C[i,:] = \sum\limits_{k = 0}^N A [i,k]*B[k,:]
\end{aligned}
\label{eq:row_wise}
\end{equation}

\begin{equation}
\begin{aligned}
C[:,j] = \sum\limits_{k = 0}^N A [:,k]*B[k,j]
\end{aligned}
\label{eq:col_wise}
\end{equation}
where $C[i,:]$ and $C[:,j]$ represent $i^{th}$ row and $j^{th}$ column, respectively, of output matrix $C$. $A$ and $B$ are the two input matrices, and $N$ is number of rows of matrix $A$ from Equation~\ref{eq:row_wise} and number of columns of matrix $B$ from Equation~\ref{eq:col_wise}.

\section{Introduction of SMASH}
\label{chap:intro:smash}

\begin{figure}[htbp]
	\centering
	\includegraphics[width=0.9\textwidth]{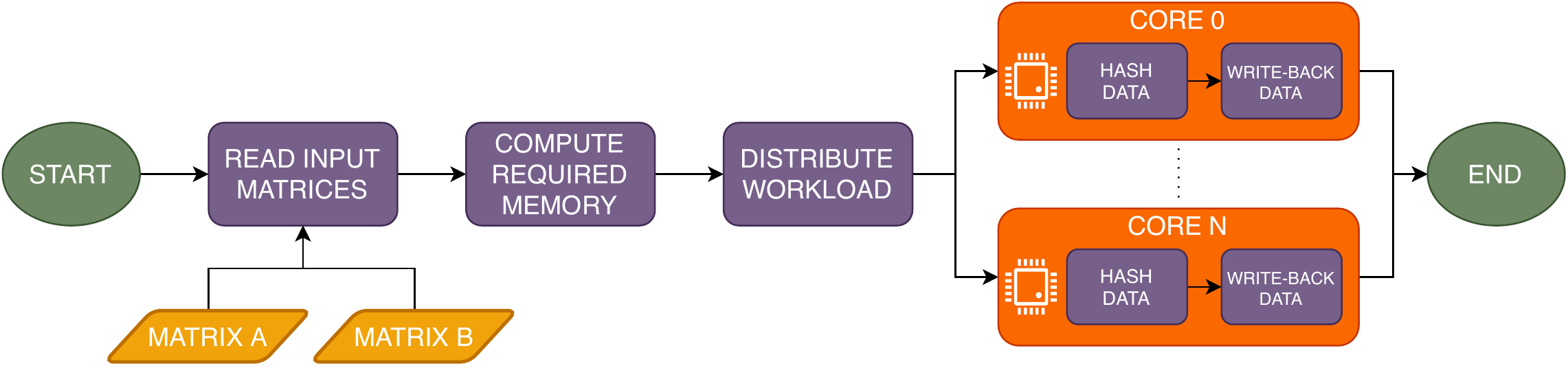}
	\caption[SMASH Overview]{\label{fig:smash_overview} SMASH Overview}
\end{figure}

In this thesis, we present Sparse Matrix Atomic Scratchpad Hashing or SMASH, a row-wise product method that uses the Compressed Sparse Row (CSR) format~\cite{smailbegovic2005sparse}. An overview of our algorithm is shown in Figure~\ref{fig:smash_overview}. The SMASH algorithm is designed to adapt to varying sparsity patterns in the input matrices. We address issues related to the row-wise product approach by designing our kernel to merge partial products using a custom implementation of a hashtable.

This thesis also discusses the unique features of a novel architecture from Intel called PIUMA, designed to speedup graph-based workloads. We further describe our implementation of \ac{SpGEMM} that exploits these features of this new architecture. Then we describe several improvements to our \ac{SpGEMM} algorithm. We discuss design decisions and their impact on the resulting algorithm. Finally, we compare our optimized \ac{SpGEMM} kernel performance on the Intel PIUMA accelerator architecture and provide an in-depth analysis of the results.

\section{Contributions}
The contributions of this thesis include:
\begin{itemize}
    \item Analysis of the problems exhibited by sparse matrix multiplication kernels.
    \item A comparison of architectures that support sparse matrix multiplications.
    \item A comparison of previous implementations of \ac{SpGEMM} kernels.
    \item An architectural overview of Intel's novel PIUMA graph accelerator.
	\item A novel \ac{SpGEMM} kernel implementation that makes the best use of the PIUMA accelerator.
\end{itemize}

%% file: tex/background.tex

\chapter{Background}
\label{chap:background}

This chapter reviews the background information on CPU and GPU architectures required to place this thesis in context. We also cover common approaches on improving the performance of \ac{SpGEMM} workloads. We include discussion on hardware designs of domain-specific accelerators for such workloads. Finally, we describe related work on \ac{SpGEMM} kernels and their implementations.

\section{CPU}
\label{chap:background:cpu}

The history of the \ac{CPU} dates back as far as 1971 when Intel introduced the first microprocessor in the market, the Intel 4004~\cite{aspray1997intel}, capable of performing 60,000 operations per second. Since then, there have been rapid advances in this field regarding clock speed and transistor technology, enabling today's AMD Ryzen CPUs to execute 2.3 teraoperations per second (a teraoperation is $10^{12}$ operations per second).

A \ac{CPU} can be defined as a computational device that, fundamentally, reads instructions from program memory and performs calculations. These instructions are fetched from the main memory of the computer (typically \ac{DRAM}) and undergo 3 stages of computations:
\begin{itemize}
	\item Fetch: Retrieve the instructions from memory. The control unit usually sends a signal through the address bus to retrieve instructions.
	\item Decode: The control unit splits the instruction into two parts, the opcode and the operands.
	\item Execute: The command represented by opcode is executed on the operand in the execute stage.
\end{itemize}

\begin{figure}[htbp]
	\centering
	\includegraphics[width=0.9\textwidth]{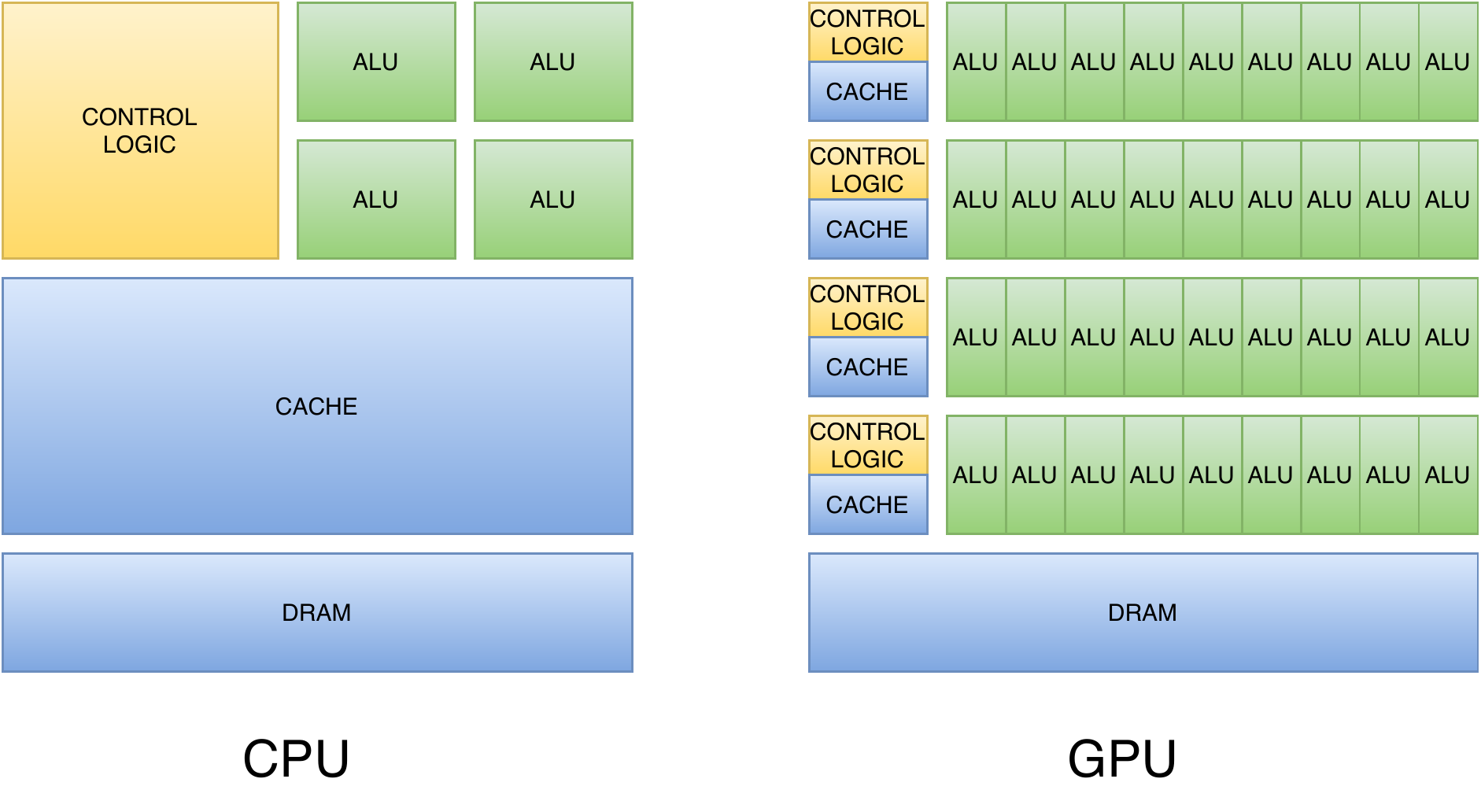}
	\caption[CPU and GPU Architectural Overview]{\label{fig:cpu_vs_gpu} CPU and GPU Architectural Overview}
\end{figure}

As far as CPU architecture is concerned, a large portion of the chip area is dedicated to control logic. With fewer cores and a large control-logic chip area, the chip real estate dedicated to control logic per core is high. In addition, every single core of the \ac{CPU} is faster than the GPU. These factors allow the \ac{CPU} to excel at certain workloads compared to the GPU.
The CPU is designed to handle a wide range of tasks efficiently but is heavily limited when running tasks concurrently.
The larger control-logic area and faster cores provide the CPUs with an advantage over the GPUs for control-dominated general-purpose workloads containing multiple conditional branches.
We compare the CPU chip area with that of GPU in Figure~\ref{fig:cpu_vs_gpu}

Given that a CPU devotes more logic to control (i.e., branch handling) and is clocked faster than the GPU, it allows the CPU to excel at executing workloads with complex single-threaded tasks, such as operating system services and database engine operations.

\section{GPU}
\label{chap:background:gpu}

\ac{GPU}s were traditionally designed to render graphics and videos to displays. They were used in appliances that need a display, like the personal computer, mobiles, and embedded systems.
Modern \ac{GPU}s are now capable of accelerating a variety of tasks that were previously executed on \ac{CPU}s. These devices led to the birth of a new field called GPU Compute, where the GPU is equipped with programmable shaders.  Today,  \ac{GPU}s commonly run a range of compute-oriented workloads, including encryption, decryption, physics simulations, pathfinding, and machine learning.

Figure~\ref{fig:cpu_vs_gpu} compares the chip area distribution between a \ac{CPU} and a \ac{GPU}. The high number of cores on a GPU allows this device to perform parallel tasks efficiently. A single core of a \ac{GPU}, compared to a \ac{CPU}, is much slower in terms of clock rate. The \ac{GPU} amortizes this slower clock speed by running thousands of tasks in parallel. Hence, workloads that possess a high degree of parallelism are better suited to run on a \ac{GPU}.

A \ac{GPU} can outperform a \ac{CPU} in many workloads that express such parallelism. Dense matrix multiplication is one such application. The high number of cores present on a single \ac{GPU} allow it to run tasks in parallel.

Although \ac{GPU}s excel at accelerating data-parallel tasks by utilizing high levels of concurrency, they do not perform well on workloads that involve control flow.  Control flow statements, such as ``if-else'' clauses, require control-flow logic in hardware to boost performance. With a large number of cores on a \ac{GPU}, the amount of chip area per core dedicated to control logic is limited. Hence, workloads that contain a significant number of conditional branches tend to perform poorly on \ac{GPU}s \cite{kerr2010modeling}.

\section{Branch and Control Flow Logic}
A computer program consists of a set of instructions. These instructions are executed in a specific order (commonly referred to as the control flow of the program).
Control flow statements (e.g., \texttt{if-else, for, while}) allow programmers to create algorithms with divergent execution sequences/paths. The decision of choosing a path is evaluated based on some conditions. The ``branch'' instruction is a special instruction that can change the execution path by altering the program counter (hence called branching).
Modern-day CPUs and GPUs overlap instruction execution in a single stream to gain speedup. This overlap of instructions is called pipelining.
Branching makes it difficult to overlap instructions because the next instruction to be executed might depend upon the previous instruction's output.

The CPU uses techniques such as branch prediction, where if the branch is predicted correctly, it incurs a little to no execution penalty, but if the branch is mispredicted, the CPU is forced to squash the contents of the pipeline and continue execution on the correct branch. 
In addition, CPUs have a deeper pipeline (with more stages) as compared to GPUs. Hence the penalty for mispredicting a branch is higher in a CPU, as more instructions will be squashed. The large control logic area enables the CPU to reduce this penalty of mispredicting branches.

A GPU executes workload in a warp (the basic unit of execution)~\cite{aamodt2018general}. A warp is a collection of GPU threads (NVIDIA GPUs contain 32 threads in a warp). GPUs do not have the resources to have each of their threads execute divergent branches simultaneously. When executing a conditional branch instruction, a single warp in the GPU computes both sides of the branch (sequentially) and discards one of them based on the correct branch path. This is referred to as predication.  Predication works for cases when each branch's size is considerably small but incurs a large penalty otherwise.
In summary, the CPU is capable of handling workloads with a large number of conditional statements, while the GPU will encounter significant slowdown for such workloads.

\section{Computational Performance}
Floating Point Operations per Second (FLOPS) is a key metric for comparing the performance of different hardware designs. This metric captures the number of floating-point operations that a device can complete in one second. Floating-point numbers can be stored in memory in different formats based on the precision required. They can be stored in \ac{HP} format (which occupies 16 bits in memory), \ac{SP} format (which occupies 32 bits), \ac{DP} format (which occupies 64 bits), or \ac{QP} format (which occupies 128 bits). \ac{SP} format stores floating-point values using 32 bits in memory, whereas the \ac{DP} format occupies 64 bits.

A modern CPU, such as Intel's Xeon 8180 Platinum Processor (from the Skylake microarchitecture family), has many  cores (the Xeon 8180 has 24 cores running at a maximum turbo frequency of $2.3$ GHz). If all cores execute AVX-512 instructions, the Xeon 8180 can reach a peak performance of $4.12$ TFLOPS~\cite{vladimirov_2020}  single-precision performance, and about $2.06$ TFLOPs of double-precision performance~\cite{vladimirov_2020} ($1\ \texttt{TFLOPS} = 10^{12}\ \texttt{FLOPS}$).

A modern GPU, such as NVIDIA's A100 GPU (based on NVIDIA Ampere architecture), has many single and double-precision cores (the A100 has $6912$ single-precision cores and $3456$ double-precision cores). The A100 can reach a peak performance values of $19.5$ TFLOPs~\cite{nvidia_2020} single-precision and $9.7$ TFLOPs \cite{nvidia_2020} double precision. 

\section{The Properties of Spatial Locality}
Spatial locality is the property that instructions and data entities tend to be stored relatively close together in an address space. Workloads exhibit high spatial locality when they request data from neighboring memory locations frequently.  
A sparse matrix multiplication kernel (\ac{SpGEMM} kernel) consists of accessing elements in random rows and columns of the input matrix (most of the rows and columns contain few non-zero elements), which results in a somewhat random memory access pattern. A large number of non-zero elements in every row can lead to high spatial locality in matrix multiplication kernels, but sparse matrices tend to have fewer non-zero elements per row, thus the resulting access-pattern of rows leads to accessing few non-zero elements scattered across the input matrix.
Hence \ac{SpGEMM} kernels exhibit low spatial locality.

This makes the \ac{SpGEMM} kernel more difficult to optimize.  In an ideal case, for efficient \ac{SpGEMM} computations, we would require a large control logic area per core (present on the CPU), along with a large number of CPU cores, to achieve the high computational performance possible on a single GPU.

\section{Sparse Matrix Storage Formats}
Matrices are generally stored in a one-dimensional linear array of contiguous memory, organized in either a row-major or column-major format. Row-major format stores subsequent elements of a row sequentially in the address space, whereas column-major stores subsequent elements of a column sequentially~\cite{thiyagalingam2003exhaustive}. While these storage formats work efficiently when working with dense matrices, they encounter performance issues when working with sparse matrices~\cite{dongarra2000sparse}. The large number of zeros present in sparse matrices cause the row-major and column-major formats to store mostly zeros in memory, thus leading to inefficient usage of valuable memory space.

The \ac{CSR} storage format stores only the non-zeros of a sparse matrix, recording the index of only the non-zero elements in each row. \ac{CSR} packs the non-zeros of each row in a single linear array (data array) and the indices corresponding to each element in another linear array (column-index array). A third array is used to track the number of elements in each row (i.e., the row-pointer array). Hence, in order to access subsequent non-zeros in a sparse matrix, one can iterate over these dense arrays of non-zeros (i.e., the data array and column-index array).
The resulting CSR format is efficient in terms of storage, as well as in terms of computation, as compared to using dense storage formats.
Similar concepts of compressed sparse storage format can be extended to storing elements of columns together. Such a matrix storage format is referred to as \ac{CSC} format.
We extensively use the \ac{CSR} storage format in our \ac{SpGEMM} kernel implementation to pack sparse matrices to fit in our on-chip memory.

%% file: tex/related_work.tex
\chapter{Related Work}
\label{chap:related_work}

Equipped with a brief introduction on the architecture of the \ac{CPU} and \ac{GPU}, and informed with an understanding of spatial locality described above, in this chapter, we discuss the class of computations that are the target of this thesis. We begin by discussing libraries that are commonly used in high-performance computing and take a deeper look into prior implementations of \ac{SpGEMM} workloads.

\section{SpGEMM on CPU}
\label{chap:background:cpu:spgemm}

The \ac{BLAS} \cite{netlib2004blas} is a standard set of libraries that provide high-performance application programming interfaces (APIs) to perform linear algebra operations. Many hardware vendors provide their own performance-tuned implementations for \ac{BLAS}, providing advantages for their own architecture. For example, Intel provides Intel \ac{MKL}~\cite{wang2014intel} for their x86 processors. Intel's implementation of these APIs exploits unique architectural features (i.e., hardware extensions) present on their \ac{CPU}s to boost their performance.
Some of these extensions include:
\begin{itemize}
	\item \ac{SSE4.2}
	\item \ac{AVX2}
	\item \ac{ABM}
	\item \ac{BMI2}
	\item \ac{FMA3}
	\item \ac{AESI}
	\item \ac{ADX}
	\item \ac{CLMUL}
	\item \ac{F16C}
\end{itemize}

Other manufacturers provide similar extensions. For the Zen architecture, 
\ac{AMD} provides the BLIS library, an optimized software implementation of the BLAS subroutines. 

In this thesis, we focus on the \ac{SpGEMM} APIs from the \ac{BLAS} library in our analysis.
In prior work, Xie et al.~\cite{xie2019ia} described an optimized \ac{SpGEMM} kernel, evaluated on both a \ac{CPU} and a \ac{GPU}. They used deep learning to train their model (called MatNet) to learn the data distribution patterns of a matrix. Their algorithm chooses the best format to represent the input data based on the MatNet model's decisions. By performing input data transformations with MatNet, they were able to accelerate a \ac{SpGEMM} kernel by $3.27\times$ over Intel's \ac{MKL} platform and $13.17\times$ speedup over \ac{AMD}'s platform.

Nagasaka et al.~\cite{nagasaka2017high} compare the performance of most publicly available implementations of \ac{SpGEMM} kernels and propose their own implementation based on hashing and heap-based algorithms. They concluded that specific implementations work better based on the pattern of non-zeros in the input data set.

\section{SpGEMM on GPU}
\label{chap:background:gpu:spgemm}

One of the largest \ac{GPU} chip manufacturers is Nvidia. Similar to the CPU vendors, Nvidia provides its own library for linear algebra kernels. In this thesis, we focus on efficient \ac{SpGEMM} kernels. Nvidia has developed its own libraries that provide \ac{SpGEMM} kernels, released as part of cuSparse and CUSP packages.
cuSparse is Nvidia's \ac{BLAS} implementation to support all sparse operations. CUSP is an open-source C++ library of generic parallel algorithms used for sparse linear algebra and graph computations on CUDA architecture GPUs.

Many other attempts have been made to optimize \ac{SpGEMM} kernels on a \ac{GPU}.
Nagasaka et al.~\cite{nagasaka2017high} present an algorithm for efficient sparse matrix multiplication on a Pascal \ac{GPU}.  Their approach uses a row-counting method (i.e., counting the number of intermediate partial-products and then grouping rows based on the number of partial-products in each row)~\cite{nagasaka2017high}. In addition to accelerating the kernel, they also try to minimize the total memory required for this operation.  They achieved a $4.3\times$ speedup on single-precision and $4.4\times$ speedup on double-precision compared to existing SpGEMM libraries. They also reduced memory usage by $14.7\%$ for single precision and $10.9\%$ for double-precision, on average.

In general, it is difficult to optimize \ac{SpGEMM} workloads on GPUs due to the random data access patterns and the large memory footprint of the intermediate data generated. We look at accelerators designed specifically for \ac{SpGEMM} in the next section.

\section{SpGEMM Accelerators}
\label{chap:background:gemm_accelerators}

Next, we focus on previous attempts made in designing domain-specific accelerators for \ac{SpGEMM} workloads. Table~\ref{table:spgemm_comparison} provides a comparison between the various accelerators and kernel implementations developed specifically for \ac{SpGEMM} workloads.

\begin{table}[ht]
	\centering 
	\begin{tabular}{p{27mm} p{22mm} p{20mm} p{55mm}}
		\hline\hline 
		\textbf{Research \newline Papers} & \textbf{SpGEMM \newline Kernel} & \textbf{Accelerator} & \textbf{Features} \\ [0.5ex]
		\hline 
		
		\ac{SpGEMM} \newline on GPU \cite{nagasaka2017high} & Outer Product & NVIDIA Pascal GPU & On-Chip shared memory merging \newline Hashtable for partial products\\
		\hline
		
		OuterSPACE \cite{pal2018outerspace} & Outer Product & OuterSPACE & Algorithm-Hardware co-design\\
		\hline
		
		ExTensor \cite{hegde2019extensor} & Inner Product & ExTensor & Hierarchical elimination of computation in the presence of sparsity\\
		\hline
		
		MatRaptor \cite{srivastava2020matraptor} & Row-wise Product & MatRaptor & New sparse storage format $ C^2SR $ \newline Hardware Sorting\\
		
		\hline
		Sunway TaihuLight \cite{chen2020optimizing} & Partitioned Outer Product & Sunway & Novel partitioning method\\
		\hline
		
		SpArch \cite{zhang2020sparch} & Outer Product & SpArch & Streaming based merger \newline Condensed matrix representation \newline Huffman tree scheduler\\
		\hline
		
		ALRESCHA \cite{asgari2020alrescha} & Inner Product & Alrescha & Data-dependent task reordering \newline Locally-dense storage format\\
		\hline
		
		Synergistic CPU-FPGA \cite{soltaniyeh2020synergistic} & Row-wise Product & CPU-FPGA & Cooperative CPU-FPGA platform \newline New intermediate representation based on communication packets\\
		\hline
		
		SIGMA \cite{qin2020sigma} & Row-wise Product & SIGMA & Novel reduction tree microarch. \newline Flexible Interconnects\\
		\hline
		
		SMASH \newline (Our approach) & Row-wise Product & PIUMA & Hashtable based on-chip merge \newline Dynamic load balancing \newline In-memory computation using PIM modules \\
		\hline
		
	\end{tabular}
	\caption{SpGEMM Accelerator Comparison} 
	\label{table:spgemm_comparison} 
\end{table}

Zhang et al. propose the SpArch accelerator~\cite{zhang2020sparch}, an accelerator designed to speed up \ac{SpGEMM} kernels. They designed a kernel using the outer-product method for matrix multiplication. The issue with outer product multiplication is the large number of intermediate partial products produced by this approach. Their work's key contribution addresses the partial product generation problem by designing a streaming-based merger into the processing pipeline, combining the multiplies with the merge stage of the partial products. This allows the partial products to be merged on-chip immediately after they are produced. 
In addition to this optimization, they also proposed a condensed matrix representation and a Huffman tree scheduler to gain further speedup. They report an average speedup of $18\times$ over Intel MKL, cuSparse, and CUSP libraries.
Although their implementation provides considerable speedup compared to other libraries, the merge-tree implementation occupies a large portion of the overall chip area. Approximately 60\% of the chip area is dedicated to the merge tree implementation, while just 1.6\% of the chip area is devoted to a multiplication array. In terms of energy, 55\% of the energy is spent on merging partial products on-chip. The extra hardware area and energy requirements devoted to partial product merging leave room for further optimization, both in terms of accelerator design and the \ac{SpGEMM} algorithm.

Qin et al.~\cite{qin2020sigma} propose SIGMA, an accelerator that tackles irregular memory accesses in \ac{SpGEMM} kernels. The fundamental block of their architecture, known as the Flexible Dot Product Engine (Flex-DPE),  consists of switchable interconnects that allow them to build a flexible network topology. Leveraging a flexible and scalable network topology allows them to keep the utilization of their processing elements high. They reported that SIGMA could obtain approximately a $3\times$ speedup compared to other state-of-the-art accelerators, including a TPU \cite{cloudtpu}, EIE~\cite{han2016eie}, SCNN~\cite{parashar2017scnn}, OuterSPACE~\cite{pal2018outerspace}, Eyeriss v2~\cite{chen2019eyeriss}, Packed Systolic~\cite{kung2019packing}, and Cambricon-X~\cite{zhang2016cambricon}.

In 2018, Pal et al. introduced their \ac{SpGEMM} accelerator called OuterSPACE \cite{pal2018outerspace}. They took a two-phase approach to implement their \ac{SpGEMM} kernel. In their first phase, called the {\em multiply phase}, they perform an outer product of the two input matrices to produce partial products. In the subsequent phase, called the {\em merge phase}, they merge these partial products to form the output matrix. Although this approach is not new, their work's novelty lies in their mapping of these phases to the OuterSPACE architecture.
The computation of an outer product when using sparse matrices causes poor data reuse and unbalanced workload distribution.
The OuterSPACE architecture is designed, keeping in mind these problems associated with \ac{SpGEMM}. With asynchronous \ac{SPMD} style worker cores coupled with memory hierarchies and shared reconfigurable caches, OuterSPACE delivered an average $7.9\times$ speedup over Intel's Math Kernel Library, $13.0\times$ over cuSPARSE, and $14.0\times$ over CUSP.

Liu et al.~\cite{liu2020systolic} introduce another accelerator that focuses on optimizing \ac{SpGEMM} kernels for mobile CNNs, using systolic arrays of Tensor Processing Elements.
ALRESCHA \cite{asgari2020alrescha} is another accelerator that differentiates itself by having two parts to its architecture; a fixed compute unit and a light-weight re-configurable engine. This allows them to adapt to input data sparsity patterns.

Many prior studies on \ac{SpGEMM} accelerator development have proposed their own sparse matrix storage format, including a condensed matrix representation in SpArch~\cite{zhang2020sparch}, the REAP intermediate representation for the CPU-FPGA accelerator~\cite{soltaniyeh2020synergistic}, $C^2SR$ for MatRaptor \cite{srivastava2020matraptor}, and Locally-Dense storage format for ALRESCHA~\cite{asgari2020alrescha}. A common issue with these schemes is that the degree of reformatting or data rearrangement needed in a sparse matrix, depending on the underlying architecture, can impact the speedup obtained by these accelerators. Memory access latency is a major bottleneck for \ac{SpGEMM} accelerators. Dividing, rearranging, and grouping data based on the sparsity patterns and accelerator architecture reduces redundant accesses and increases data locality. Though these novel formats boost performance, they incur associated conversion costs since the sparse input matrices are typically stored in \ac{CSR}, \ac{CSC}, or \ac{ELL} storage formats. These costs are either in terms of additional hardware requirements or performance degradation or both. The design of any storage format should consider whether the benefits provided can amortize the data transformation cost.
Considering previous efforts to speed up sparse workloads and considering the problems faced by these accelerators, we propose our own \ac{SpGEMM} kernel implementation developed on a general hardware accelerator architecture. 

%% file: tex/architecture.tex

\chapter{PIUMA Architecture and Simulator}
\label{chap:piuma}

The \ac{PIUMA} is being developed by Intel \cite{aananthakrishnan2020piuma}, as part of DARPA's {\em Hierarchical, Identify, Verify, Exploit (HIVE)} program~\cite{jacobs}. The \ac{HIVE} project recognizes the challenges involved with graph analytics and aims to achieve a 1,000$\times$ performance/Watt improvement over the previous state-of-the-art system~\cite{mccreary_2020}.
The \ac{PIUMA} system is a scalable systems architecture designed to accelerate graph-based applications. The system is designed to handle
sparsity, bearing in mind the highly-random data access patterns present in graph workloads.

\section{PIUMA Architecture}

This section provides a detailed description of Intel's \ac{PIUMA} machine, exploring some of the key features of this architecture that are exploited in our implementation of the \ac{SpGEMM} kernel. We highlight some of the problems faced in \ac{SpGEMM} and focus on PIUMA's components that help tackle these challenges. Primarily, we focus on the following architectural features of PIUMA~\cite{aananthakrishnan2020piuma}:
\begin{enumerate}
    \item PIUMA Cores
    \begin{enumerate}
        \item Multi-Threaded Core (MTC) and
        \item Single-Threaded Core (STC)
    \end{enumerate}
    \item Offload Engines (OE)
    \begin{enumerate}
        \item DMA engine and
        \item Collective engine
    \end{enumerate}
    \item Global Address Space
    \item Network
\end{enumerate}

\subsection{PIUMA Cores}
The \ac{PIUMA} cores form the fundamental computational unit of this architecture.  They are designed to exploit the inherent parallelism exhibited by graph-based workloads.
In general, graph workloads are more memory intensive than compute-intensive workload~\cite{eyerman2018many}. A key principle in \ac{PIUMA} is to provide a high degree of parallelism to hide memory latency. \ac{PIUMA}'s large number of threads are capable of keeping many memory requests in flight.
The \ac{PIUMA} cores can be classified into two types:
\begin{enumerate}
    \item The \ac{MTC} and
    \item The \ac{STC}
\end{enumerate}
We take a closer look into the architectural layout of each of these cores.

\subsubsection{The PIUMA Multi-threaded Core}
A multi-threaded core consists of one pipeline each~\cite{aananthakrishnan2020piuma}. The \ac{MTC} can issue at most one instruction per cycle, providing for an energy-efficient design~\cite{aananthakrishnan2020piuma}. In addition, the \ac{MTC} also has an associated register file \ac{RF}. Each \ac{RF} stores the context of up to 16 threads simultaneously, allowing a single \ac{MTC} to be shared by up to  16 threads. Each thread represents a single stream of executions. The \ac{MTC} resources are shared across these 16 threads, utilizing a round-robin resource allocation routine.
When a stage of the MTC's pipeline stalls, a new thread is swapped to execute, hiding latency and keeping the pipeline stages full.

\subsubsection{Single-Threaded Core}
Unlike the \ac{MTC}, the \ac{STC} comprises a single thread of execution (executing a single stream of instructions).
Since a single instruction stream is being executed every cycle (as opposed to the round-robin scheduling of \ac{MTC}s), a higher priority is given to this single thread, making it capable of handling performance-sensitive tasks~\cite{aananthakrishnan2020piuma}.
The \ac{STC}s are in-order blocking (on misses), cores designed to lower power consumption as compared to the out-of-order pipelines~\cite{aananthakrishnan2020piuma}.
The primary purpose of \ac{STC}s is to perform memory and thread management tasks.

Both the single-threaded and multi-threaded cores are equipped with L1 instruction caches and L1 data caches. 
Graph applications are known for their irregular data access patterns~\cite{eyerman2018many}. The resulting irregular memory access patterns lead to low L1, L2, and L3 cache utilization due to low spatial locality~\cite{kelly_2018}.
As graph workloads exhibit poor locality, in \ac{PIUMA} architecture, no higher cache levels are included to save on power consumption~\cite{aananthakrishnan2020piuma}. All caches throughout the \ac{PIUMA} system are non-coherent.  Although cache coherency is helpful to maintain data uniformity, it is associated with large overheads. Caches can be made coherent using protocols such as $MSI$, $MESI$, $MOSI$, and $MOESI$ (Modified, Owned, Exclusive, Shared, Invalid)~\cite{hackenberg2009comparing}.
These protocols can cause the executing pipeline to stall for thousands of cycles to fetch the new data value from the main memory or neighboring caches.
The \ac{PIUMA} architecture does not provide cache coherency. It thus becomes the responsibility of the programmer to avoid modifying shared data and flush caches as required~\cite{srinivasan2016dynamic}.  In addition, prefetching is disabled to limit power consumption~\cite{aananthakrishnan2020piuma}.

\begin{figure}[htbp]
	\centering
	\includegraphics[width=1.0\textwidth]{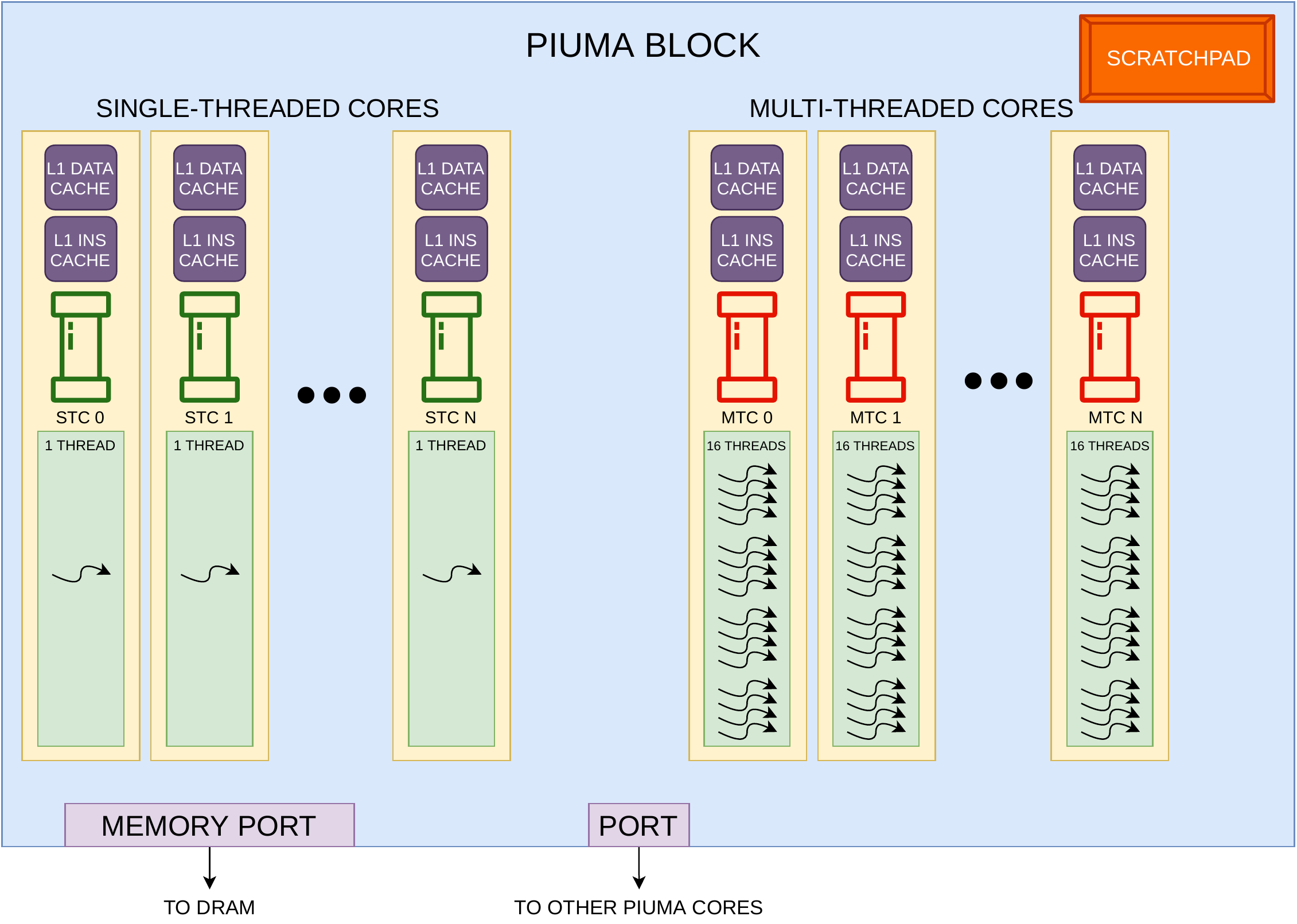}
	\caption[PIUMA Core]{\label{fig:piuma_block} A single PIUMA block.}
\end{figure}

Multiple \ac{MTC}s and \ac{STC}s are grouped in a block (see Figure~\ref{fig:piuma_block}). Each block consists of low latency, user-accessible storage, scratchpad (SPAD) memory. Programmers can use this shared storage for storing data with high temporal locality~\cite{spmv_piuma}.

\subsection{Offload Engines}

Traditionally, following the Von Neumann architecture~\cite{SHIPLEY2003545}, programs and data are commonly stored in memory and fetched by the processor for execution. This architecture will be limited by the throughput of the memory channels, commonly known as the Von Neumann bottleneck~\cite{10.1007/978-3-540-85373-2_8}. In some novel architectures~\cite{lin2007multiplier, nair2020defect,peng2019dnn+}, this limitation was overcome by introducing \ac{C-RAM} technology (i.e., processors in memory). \ac{C-RAM} is similar to \ac{DRAM} but with a vector processing element embedded on the same chip as the memory. This enabled \ac{C-RAM} chips to service instructions other than a simple load or store and perform operations such  as scatter and gather.

The \ac{PIUMA} architecture provides Offload Engines (OE) to support the \ac{PIUMA} cores in memory-related operations. This allows selected \ac{SIMD} instructions to be executed in memory. Some examples of such \ac{SIMD} instructions supported by \ac{PIUMA} include:

\begin{enumerate}
    \item $Copy$: Copy a chunk of data from one section of memory to another.
    \item $Strided\ Copy$: Similar to a copy, but every $n^{th}$ element is copied.
    \item $Gather$: Read an array of data and compute its sum.
    \item $Scatter$: Broadcast a single value to multiple locations in memory.
\end{enumerate}
We discuss two offload engines that we use to speed up our \ac{SpGEMM} algorithm.

\subsubsection{DMA Engine}
Data movement forms an integral part of graph algorithms. One of the major bottlenecks in graph applications is memory throughput~\cite{eyerman2018many}. The \ac{DMA} Engine reduces the workload on \ac{PIUMA} cores by executing the memory operations (i.e.,  $loads$ and $stores$).
A single $copy$ instruction or $gather/scatter$ instruction can replace thousands of $load$ and $store$ instructions issued by the cores.
The DMA engine carries out the underlying task of copying multiple bytes of data or broadcasting data to multiple memory locations.
All instructions issued to the \ac{DMA} engine run in the background (non-blocking), freeing the core to execute other instructions.

\subsubsection{Collective Engine}
One important element of many parallel algorithms is the required synchronization. The \ac{PIUMA} architecture is equipped with a collective engine that provides system-wide barriers and reduction operations~\cite{aananthakrishnan2020piuma}.

One of the key features presented by the \ac{PIUMA} architecture is the ability to offload instructions over the fabric to remote cores. Threads can wrap their instructions in network packets and forward them to remote threads for execution. These instructions are called remote or network instructions.

When a thread intends to update data present in remote memory (memory physically connected to neighboring cores), it can send remote instructions to be executed by another thread, local to that memory. Thus, instead of streaming data over the network, we stream instruction packets to the core that is physically connected to that memory chunk. Network instructions can be helpful in two ways:
\begin{enumerate}
    \item forwarding instructions to threads that have low latency access to data can improve performance, and
    \item distributing the workload among threads can improve workload balance.
\end{enumerate}

 Remote atomic instruction is one such networked instruction that allows atomic instructions to be executed by \ac{PIUMA} cores in memory that is remote. We make use of remote atomics in our algorithm to update the partial products in our hash table.

\begin{figure}[htbp]
	\centering
	\includegraphics[width=1.0\textwidth]{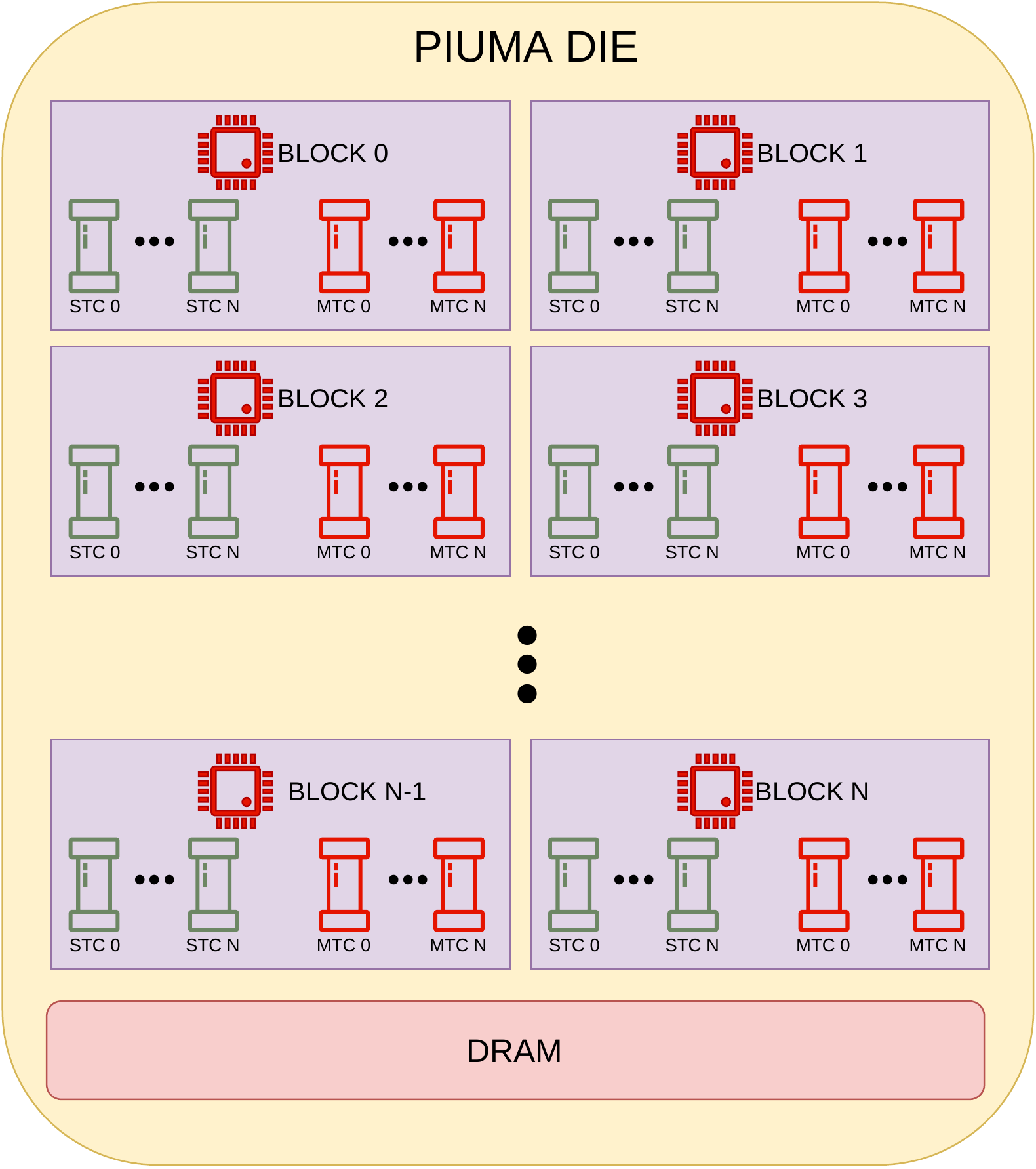}
	\caption[PIUMA Die]{\label{fig:piuma_die} A single PIUMA die.}
\end{figure}

\subsection{Global Address Space}
\ac{DGAS} is a model in which the memory address space is logically divided, and sections of this address space are local to each computing thread. \ac{DGAS} allows using an SPMD style of programming~\cite{yelick2007productivity}, while supporting data addressing semantics similar to a shared memory system.

The \ac{PIUMA} architecture is built using \ac{DGAS}, allowing data present on any core to be accessible by any other \ac{PIUMA} core/thread.  This allows the programmer to worry less about the scope of memory access pointers and focus instead on parallelizing the workload's execution.
Each \ac{PIUMA} thread has an affinity to specific sections of the \ac{DGAS}. Despite each thread having access to the entire address space, local memory accesses will experience lower latency than remote accesses. Thus, data stored in the \ac{DGAS} partition belonging to a thread is said to have an affinity to that thread.  We use \ac{DGAS} when designing our sparse matrix algorithm to broadcast sections of the input matrix from the first core to all other cores involved in computing the workload. This will transfer the elements to each thread's local memory, as described in Section~\ref{chap:smash} in detail. 

The \ac{PIUMA} system consists of Address Translation Tables (ATT). ATT contains reconfigurable rules to translate application memory addresses to physical memory locations, enabling us to rearrange the address space as needed by the application~\cite{aananthakrishnan2020piuma}.

In addition, the \ac{PIUMA} memory controllers are redesigned to support native 8-byte accesses~\cite{aananthakrishnan2020piuma}. Instead of a full cache line fetch, the memory controllers can selectively fetch 8-byte words, reducing redundant memory fetches.

\subsection{Network}
The \ac{PIUMA} network connects blocks (groups of \ac{MTC}s and \ac{STC}s) together and forwards memory requests to remote memory controllers.
The \ac{PIUMA} system is configured in a HyperX topology to achieve high bandwidth and low latency \cite{aananthakrishnan2020piuma,ahn2009hyperx}. This allows the network to have a high radix and a low diameter.
The higher-level links are optical to sustain high-bandwidth at low power consumption levels.

\begin{figure}[htbp]
	\centering
	\includegraphics[width=0.75\textwidth]{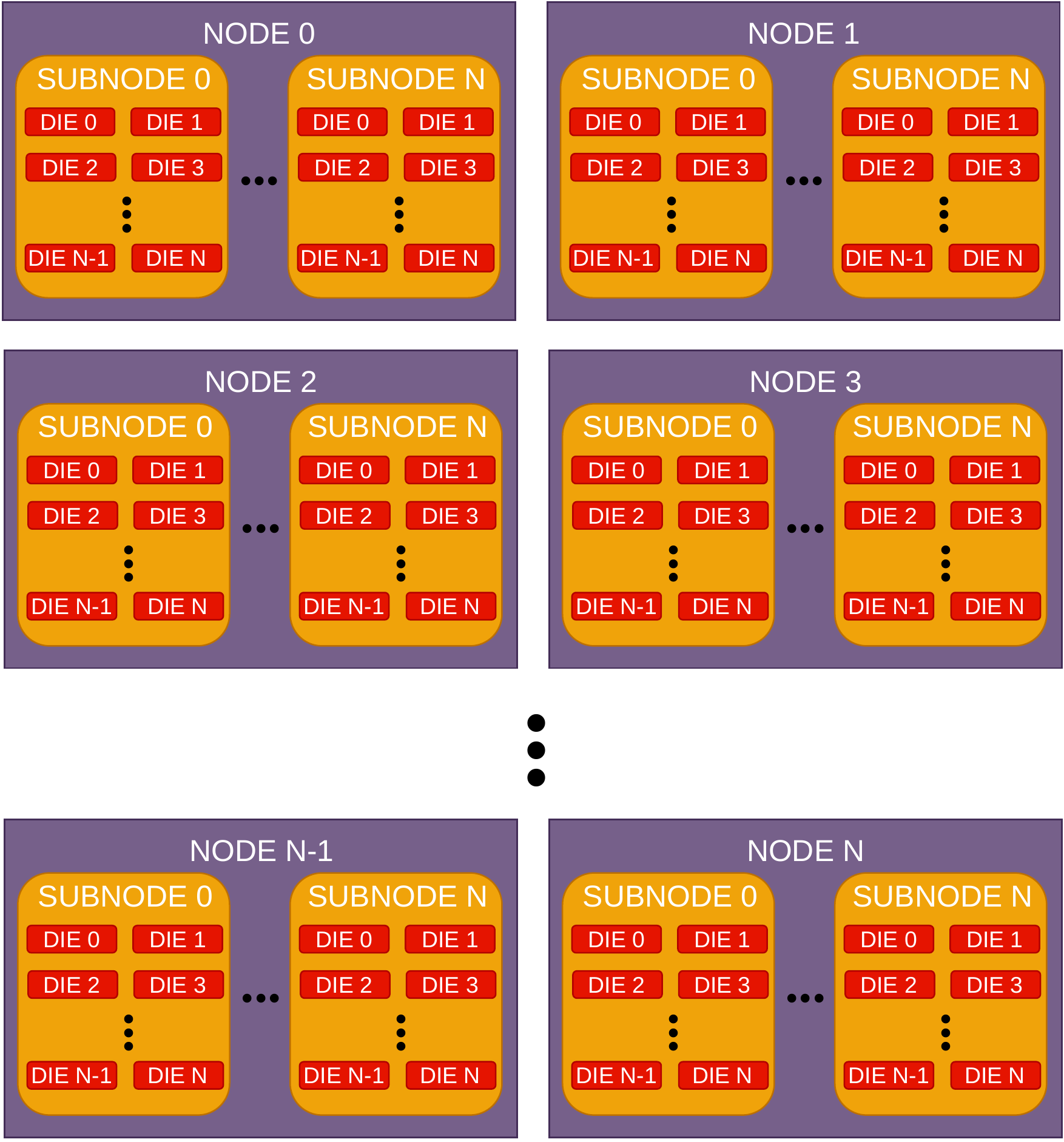}
	\caption[PIUMA System]{\label{fig:piuma_system} The PIUMA system.}
\end{figure}

A single \ac{PIUMA} block consists of both the \ac{STC}s and \ac{MTC}s. Each \ac{STC} and \ac{MTC} is accompanied by an L1 instruction cache and an L1 data cache. Each block is accompanied by local high-bandwidth Scratchpad memory (see Figure~\ref{fig:piuma_block}).
Multiple \ac{PIUMA} blocks are laid out together to form a die, as shown in Figure~\ref{fig:piuma_die}).
Figure~\ref{fig:piuma_system} presents the layout of an entire \ac{PIUMA} system at node, subnode, and die level.

The \ac{PIUMA} system is a novel approach to tackle the problems present in graph-based workloads. The simple, in-order, multi-threaded cores and non-coherent caches aid in hiding latency of random memory accesses present in graph applications. The offload engines, including the DMA engine and collective engine, work alongside the \ac{PIUMA} cores to help with memory operations and synchronization. The distributed shared global address space makes it convenient for programmers to implement distributed kernels. Finally, the HyperX topology of the \ac{PIUMA} network with optical interconnects delivers a design that can scale out to multiple nodes, allowing computation of graphs with trillions of vertices~\cite{aananthakrishnan2020piuma}.

\section{Simulation Methodology}

The evaluation of new computer architecture features is commonly evaluated pre-silicon using a simulator \cite{akram2019survey}.
Simulators are software models used to simulate the behavior of various design features. In contrast to simulation models, analytical models are statistical models that can mathematically evaluate elements of a computer architecture. Owing to the complexity of today's computer architectures and the large number of configurable parameters, analytical models tend to produce inaccurate results, hence are not suitable for evaluating computer architectures~\cite{akram2019survey}.

\begin{figure}[htbp]
	\centering
	\includegraphics[width=0.75\textwidth]{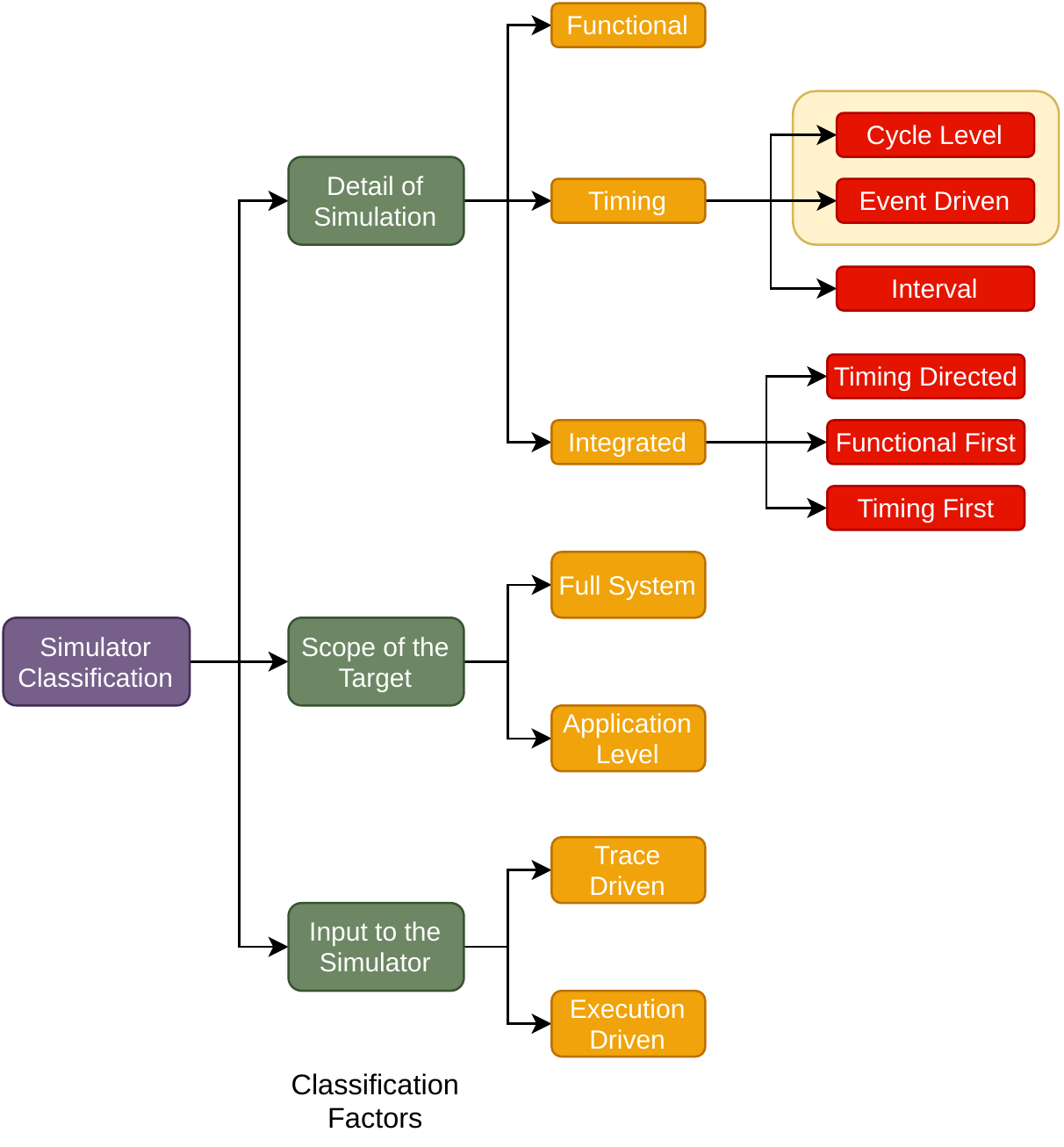}
	\caption[PIUMA System]{\label{fig:sim_classify} execution-driven simulator.} 
\end{figure}

The design cycle and silicon fabrication process required to produce physical hardware is both time-consuming and can incur  high cost. Simulators allow architects to make design decisions before the hardware pre-silicon, thus lowering the hardware development costs~\cite{akram2019survey}, making them an integral part of today's hardware design process. 

\subsection{Simulator Classification}
For our simulation of \ac{PIUMA} architecture, we use the Sniper simulator \cite{heirman2012sniper}. To begin, we first review different classes of simulators to better understand the advantages of the Sniper simulator over others choices.  Simulators can be grouped into various classes based on a range of factors, including the level of detail of the simulation, the scope of the target (the system that is being simulated), and input that drives the simulator.
Figure~\ref{fig:sim_classify} provides an overview of the various classes of simulator~\cite{akram2019survey}. We briefly describe each class and discuss the advantages/disadvantages provided by each.

Our classification begin by consider the level detail of simulation:
\begin{itemize}
    \item \textbf{Functional Simulators}: Functional simulators focus on the functionality of the modeled architecture. They provide for emulation of the target ISA, but they do not implement the underlying microarchitecture of the target system, making them faster than other simulators. 
    \item \textbf{Cycle-level Simulators}: Cycle-level simulators model the operations of a processor cycle-by-cycle. Cycle-level simulators are relatively slower than functional simulators and consume a significant amount of memory resources.
    \item \textbf{Event-driven Simulators}: Event-driven simulators target events instead of cycles. Simulations steps through time to arrive at events when they are scheduled~\cite{martin2005multifacet}, thus saving time by not simulating cycles that are not scheduled.
    \item \textbf{Interval Simulators}: While cycle-level simulators are accurate, they  are very slow. Event-driven simulators are fast, but compromise accuracy. Interval simulators strike a balance between speed and accuracy~\cite{genbrugge2010interval}. Interval simulators work on the fact that instructions flow through a pipeline can be broken down into sets of intervals, based on miss events (miss events include branch mispredictions and cache misses). A distinct branch misprediction simulator and cache miss simulator can then be used to accurately evaluate every interval's performance.
    \item \textbf{Integrated Timing-directed Simulators}: Functional simulators are often integrated with timing simulators. In a timing-directed simulator, the functional-simulator records the architectural state (register and memory) of the processor and forwards it to the timing simulator. The timing simulator then uses these values to perform corresponding computations. There exists heavy communication between the functional simulator and the timing simulator.
    \item \textbf{Integrated Functional-first Simulators}: In a functional-first simulator, a functional-simulator leads the timing simulator. The functional simulator generates an instruction trace and forwards it to the timing simulator. Functional-first simulators are similar to trace-driven simulators, with a distinguishing factor being the traces are generated by the functional simulator and forwarded to the timing simulator immediately while trace-driven simulators store traces on file after generation.
    \item \textbf{Integrated Timing-first Simulators}: In a timing-first simulator, the timing simulator leads the functional simulator. The timing simulators simulate the microarchitecture at cycle-level and then use functional simulators for verification purposes. In cases where the timing simulator results do not match the functional simulator, the timing simulator flushes its pipeline and restarts from the fetch cycle for that instruction.
    \end{itemize}
Classification based on the scope of the target architecture is as follows:
    \begin{itemize}
    \item \textbf{Full System Simulators}: Full system simulators support booting the entire operating system (OS) and run target applications in that OS. Full system simulators are significantly slower and are generally used to simulate I/O devices.
    \item \textbf{Application-level Simulators}: As opposed to full system simulators, application-level simulators only simulate target applications. Since they do not suffer the high overhead of simulating the operating system or system calls.  This class of simulator is significantly faster.
\end{itemize}
Classification based on input to the simulator is as follows:
\begin{itemize}
    \item \textbf{Trace-driven Simulators}: Trace-driven simulators use trace files as input. Trace files are pre-recorded streams of instructions from a previous run of the application. Trace files are usually stored in a file system and occupy a large amount of space, which in some cases, becomes a bottleneck for simulation~\cite{john20028}\cite{eeckhout2010computer}.
    \item \textbf{Execution-driven Simulators}: In contrast to trace-driven simulators, execution-driven simulators use application binaries as input instead of trace-files. These application binaries are significantly smaller in size as compared to trace files. Execution-driven simulators can simulate misspeculated instructions, unlike trace-driven simulators.
\end{itemize}

\subsection{Sniper Simulator}

Sniper is a parallel, multi-core x86 architecture simulator~\cite{heirman2012sniper}. For functional simulation, Sniper uses the Graphite simulator, which is based on the Pin tool \cite{miller2010graphite}. Sniper is classified as an Application-level simulator, making it faster than full system simulators.

Sniper is an interval-based simulator. Instead of simulating each instruction, Sniper breaks down the stream of instructions into discrete sets called intervals. These intervals are based on events, such as cache misses and branch predictions. Hence, the Sniper simulator is significantly faster owing to it's ability to '\textit{jump}' between miss events~\cite{heirman2012sniper}. A special branch predictor simulator coupled with a memory system simulator can then be used to evaluate miss events. The metrics of these simulators are then compiled with the analytical model's findings to estimate the duration of every interval~\cite{akram2019survey}.

In this thesis, to evaluate performance of our proposed scheme,  we use a modified implementation of the Sniper simulator.  An interval-based simulator, like Sniper, simulates processors at a higher level of abstraction. Using such a simulator allows us to simulate multi-core processors efficiently (several million instructions per second), as compared to a detailed cycle-accurate simulator. 
Although cycle-level simulators, such as Gem5 \cite{binkert2011gem5}, are more accurate than high-level simulators, they tend to be significantly slower, limiting our options to simulate a range of hardware configurations~\cite{carlson2014aeohmcm}.
To record these observations, we use a host machine based on the machine configuration described in Table~\ref{table:host_specs}.
Our target architecture simulator configuration is as shown in Table \ref{table:simulator_config}.

\begin{table}[ht]
	\centering 
	\begin{tabular}{l c} 
		\hline\hline 
		\textbf{Item} & \textbf{Description} \\ [0.5ex] 
		\hline 
		Manufacturer & Intel Corporation\\ 
		System Details & Intel Server Board S2600TP Family \\
		CPU & Intel Xeon CPU E5-2699A\\
		Threads per core & 2 \\
		Total cores per CPU & 18 \\
		Number of Sockets & 2 \\
		Total threads in node & 72 \\
		Max CPU Frequency & 3.4 GHz \\
		RAM & 256 GB \\
		OS & CentOS Linux release 7.7.1908 (Core) \\ [1ex] 
		\hline 
	\end{tabular}
	\caption{Simulator host machine specifications.} 
	\label{table:host_specs} 
\end{table}

\begin{table}[hbt!]
	\centering 
	\begin{tabular}{l l l l} 
		\hline\hline 
		\textbf{Scope} & \textbf{Configuration} & \textbf{Value} & \textbf{Description} \\ [0.5ex] 
		\hline 
		
		\multirow{5}{*}{Machine Global} & \multicolumn{1}{l}{Rack Count} & \multicolumn{1}{l}{1} & \multicolumn{1}{p{5cm}}{Number of racks in the System} \\\cline{2-4}
		& \multicolumn{1}{l}{Board Count} & \multicolumn{1}{l}{1} & \multicolumn{1}{p{5cm}}{Number of socket units per rack} \\\cline{2-4}
		& \multicolumn{1}{l}{Socket Count} & \multicolumn{1}{l}{1} & \multicolumn{1}{p{5cm}}{Number of sockets} \\\cline{2-4}
		& \multicolumn{1}{l}{Die Count} & \multicolumn{1}{l}{1} & \multicolumn{1}{p{5cm}}{Number of dies} \\\cline{2-4}
		& \multicolumn{1}{l}{Core Count} & \multicolumn{1}{l}{Varying (1 to 8)} & \multicolumn{1}{p{5cm}}{Number of cores per die} \\\hline
                         
		\multirow{2}{*}{Socket Global} & \multicolumn{1}{l}{DRAM Count} & \multicolumn{1}{l}{1} & \multicolumn{1}{p{5cm}}{Number of MCs to external DRAM banks per socket} \\\cline{2-4}
		& \multicolumn{1}{l}{DRAM Size} & \multicolumn{1}{l}{unlimited} & \multicolumn{1}{p{5cm}}{Size in MB per MC for external DRAM} \\\hline

		\multirow{8}{*}{Core Global} & \multicolumn{1}{l}{STC Count} & \multicolumn{1}{l}{2} & \multicolumn{1}{p{5cm}}{Number of STCs per core} \\\cline{2-4}
		& \multicolumn{1}{l}{MTC Count} & \multicolumn{1}{l}{4} & \multicolumn{1}{p{5cm}}{Number of MTCs per core} \\\cline{2-4}
		& \multicolumn{1}{l}{Core SPAD Count} & \multicolumn{1}{l}{1} & \multicolumn{1}{p{5cm}}{Total logical Scratchpad enteries per block} \\\cline{2-4}
		& \multicolumn{1}{l}{Core SPAD Size} & \multicolumn{1}{l}{4,096} & \multicolumn{1}{p{5cm}}{Size of each scratchpad in KB} \\\cline{2-4}

		& \multicolumn{1}{l}{Cache Size} & \multicolumn{1}{l}{16} & \multicolumn{1}{p{5cm}}{Size in KB for Cache Module} \\\cline{2-4}
		& \multicolumn{1}{l}{Cache Assoc.} & \multicolumn{1}{l}{4} & \multicolumn{1}{p{5cm}}{Associativity of the cache} \\\cline{2-4}
		& \multicolumn{1}{l}{Cache Line Size} & \multicolumn{1}{l}{64} & \multicolumn{1}{p{5cm}}{Line size in bytes of cache module} \\\cline{2-4}
		& \multicolumn{1}{l}{Cache Policy} & \multicolumn{1}{l}{wb-wa} & \multicolumn{1}{p{5cm}}{Replacement policies of the cache (Write-back, Write Allocate)} \\\hline

		\hline 
	\end{tabular}
	\caption{Simulator target configuration for PIUMA architecture} 
	\label{table:simulator_config} 
\end{table}

%% file: tex/smash.tex

\chapter{SMASH Kernels}
\label{chap:smash}

We have reviewed the problems present in prior sparse matrix multiplication algorithms.  We have  also described the \ac{PIUMA} architecture that we will target in this work.  In this chapter, we will present a new \ac{SpGEMM} kernel that can fully exploit the features of the \ac{PIUMA} machine.

One of the key design points for our \ac{SpGEMM} kernel implementation will be to choose between the four general matrix-multiplication approaches (presented in Figure~\ref{fig:matrix_multiplication}).
The inner product approach for sparse matrix multiplication faces issues due to the slow index-matching process, in addition to exhibiting poor input data reuse \cite{pal2018outerspace}.
The outer-product approach generates a large number of intermediate partial products. These partial products have to be buffered somewhere for later merging. The high on-chip memory requirements of the outer-product approach make it unsuitable for multiplying extremely sparse matrices.

We introduce our novel implementation of the SpGEMM kernel based on a row-wise product method called \textit{Sparse Matrix Atomic Scratchpad Hashing} or (SMASH). The row-wise product method is beneficial given its high input reuse behavior~\cite{qin2020sigma}. It gives us the ability to perform a minimum number of input matrix reads while maintaining low on-chip memory usage.

In this thesis, we present 3 different versions of our SMASH kernel. Each version improves the efficiency of a different section of our \ac{SMASH} implementation.

We adopt an iterative improvement approach to identify bottlenecks in the current version and modify our algorithm to mitigate them in the next version.
The following sections will describe in detail the 3 different versions of the \ac{SMASH} kernel.

\section{SMASH Version 1: Atomic Hashing}

A row-wise product method multiples each element of the first input matrix with an entire row of the second input matrix to generate a partial product for the output matrix. These partial products are then merged to form output matrix elements.

Each row of the first input matrix, when multiplied by their corresponding rows from the second input matrix, will generate a series of partial product matrices, as seen from Equation~\ref{eq:row_wise}.
This is one of the disadvantages of using a row-wise product method. The intermediate results  (partial products) generated need to be stored in the main memory and refetched to be merged back into the output matrix.  We overcome this obstacle with our first implementation of the \ac{SMASH} kernel by using atomic hashing. Instead of writing the partial products back to memory, we implement a streaming mechanism to merge them on-chip, thus avoiding redundant writes to \ac{DRAM}.

\begin{figure}[htbp]
	\centering
	\includegraphics[width=0.9\textwidth]{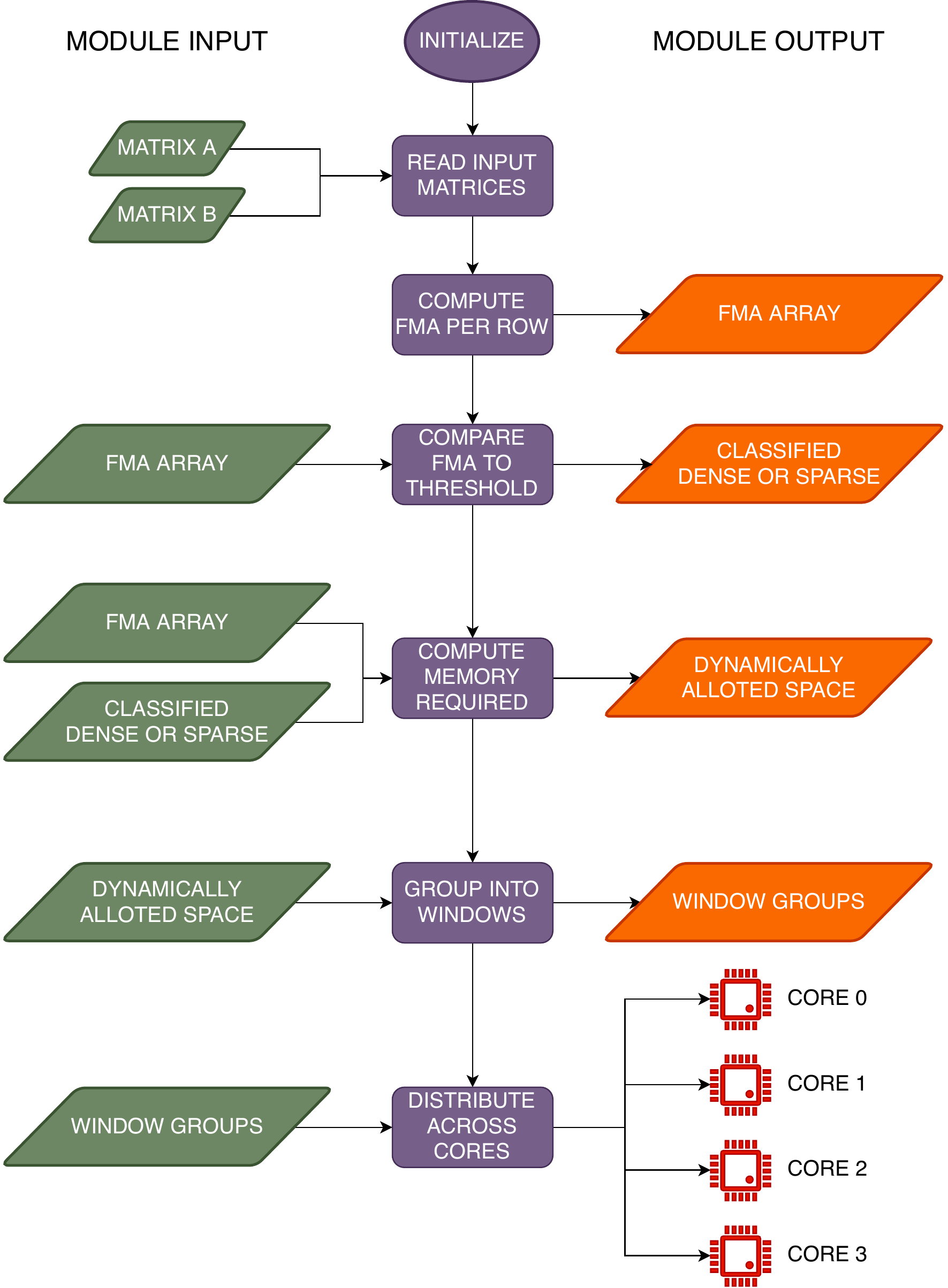}
	\caption[Window Distribution Algorithm]{\label{fig:window_distro} Window Distribution Algorithm.}
\end{figure}

The \ac{SMASH} implementation can be distinctively divided into three different phases.
\begin{enumerate}
	\item the window distribution phase,
	\item the hashing phase and 
	\item the write-back phase.
\end{enumerate}

Each phase's completion is accompanied by a synchronization barrier that spans the entire \ac{PIUMA} system.

\subsection{Window Distribution Phase}

Our window distribution phase begins with reading both input matrices. We store the input data matrices in the \ac{CSR} format. This helps us in two key respects:
\begin{enumerate}
	\item The \begin{math} row\_pointer \end{math} array of the \ac{CSR} format allows us to compute the amount of memory required to allocate the output matrix. This computation is inspired by Gustafson's algorithm for \ac{SpGEMM} \cite{gustavson1978two, filev2010extended, georgieva2009gustafson, filev2010extended, krishnapuram1999note, seo2012parallel}.
	\item The element access pattern of the row-wise product method involves obtaining rows of input matrices. Thus, the data arrangement in the \ac{CSR} storage format improves the spatial locality pattern of our solution.
\end{enumerate}

After reading the input matrix arrays in \ac{CSR} format, we compute the required amount of memory needed to store the output matrix by counting the total \ac{FMA} operations per row. To accomplish this, we use Gustafson's two-step algorithm~\cite{gustavson1978two}. The computation of the total number of \ac{FMA}s per row has a computational complexity of $\mathcal{O}(n)$, where $n$ is one of the dimensions of the input matrix.

Once an array of \ac{FMA}s is generated, we decide, on a row-by-row basis, if each row should be computed as a dense row or a sparse row. The decision is made by using a threshold value specifying the maximum number of elements that need to be present in a sparse row.

Next, we group multiple rows in a single window to be computed by one \ac{PIUMA} block. The size of a window is a function of the \ac{SPAD} size.
Sections of input matrices are then packaged and shipped to individual blocks in network packets using \ac{PIUMA}'s global address space feature. This data is then stored in the block's \ac{DRAM}, ready to be processed.

Every individual \ac{PIUMA} block processes its own window independently, regardless of the status of other windows. This allows us to schedule windows to blocks in random order and oversubscribe windows to blocks (Blocks with windows containing largely sparse rows can be oversubscribed as they will end up completing before other windows).
Details of the window set up and distribution can be seen in the Algorithm \ref{alg:setup}.

\begin{algorithm}[htbp]
\label{alg:setup}
\linespread{0.5}\selectfont
\SetAlgoLined
\caption{SMASH SETUP}
\DontPrintSemicolon
\KwIn{Matrix to be multiplied: $mat\_A$ (Stored on DRAM)}
\KwIn{Matrix to be multiplied: $mat\_B$ (Stored on DRAM)}
\KwOut{Final output matrix: $mat\_C = (mat\_A \times mat\_B)$ (Stored on DRAM)}
\tcp*[h]{Read $mat\_A$ in CSC format}\;
$A\_col\_ptr \leftarrow$ Array of column pointers of matrix A in CSC format\;
$A\_row\_idx \leftarrow$ Array of row indices of matrix A in CSC format\;
$A\_data \leftarrow$ Array of data values of matrix A in CSC format\;
$B\_row\_ptr \leftarrow$ Array of row pointers of matrix B in CSR format\;
$B\_col\_idx \leftarrow$ Array of col indices of matrix A in CSR format\;
$B\_data \leftarrow$ Array of data values of matrix A in CSR format\;
$A\_col\_ptr\_copy\_1 \leftarrow$ First copy of column pointer array of A\;
$A\_col\_ptr\_copy\_2 \leftarrow$ Second copy of column pointer array of A\;
$hash\_size \leftarrow SPAD\_SIZE$\;
$matrix\_size \leftarrow 2^{17}$ \tcp*[h]{Number of rows or columns in matrix A}\;
$window\_size$ \tcp*[h]{Computed dynamically for each window}\;
$element\_size \leftarrow$ Size\ of\ one\ element\;
\textcolor{red}{Launch tasks on all threads}\;
$tid \leftarrow$ Unique thread ID for every thread\;
$hash\_shift \leftarrow log_2(\frac{Total\ bins\ in\ Window}{Total\ bins\ in\ SPAD})$\;
EMPTY $\leftarrow -1$ \tcp*[h]{A unique flag}\;

\For{$w \leftarrow Total\ Windows$} {
\tcp*[h]{HASHING PHASE}\;
\barrier\;
\tcp*[h]{WRITEBACK PHASE}\;
\barrier\;
}
\end{algorithm}

\subsection{Hashing Phase}

In the hashing phase, a global hashtable is created in the \ac{SPAD}. A single row is allocated to one thread of each \ac{MTC} in a round-robin fashion. Each element of the row from the first matrix is multiplied with an entire corresponding row of the second matrix. This leads to the creation of partial products. These partial products are hashed into the \ac{SPAD} using bit-shift hashing.

\begin{figure}[htbp]
	\centering
	\includegraphics[width=0.3\textwidth]{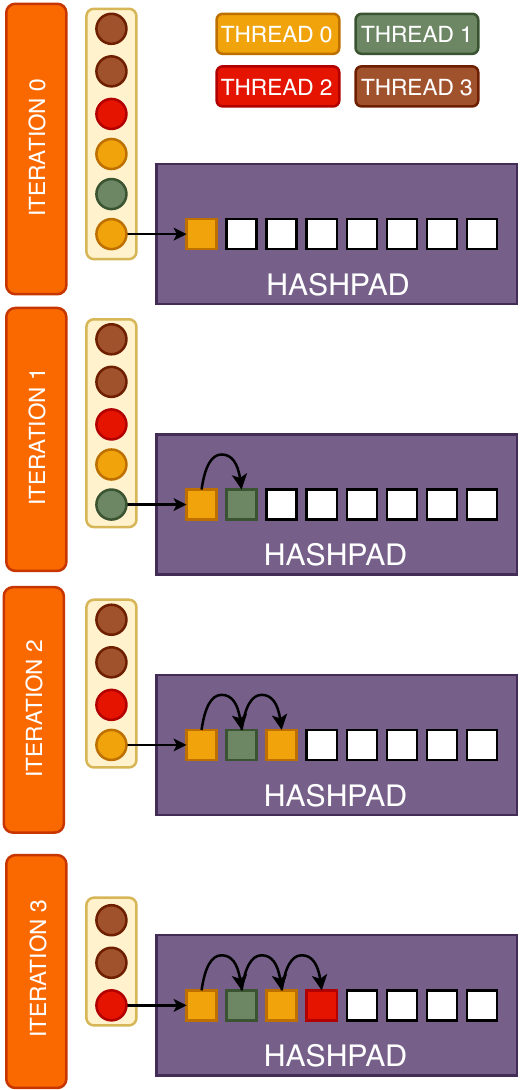}
	\caption[Collision Resolution]{\label{fig:collision_resolution_upper} Collision Resolution.}
\end{figure}

To hash using bit-shifts, we ignore the lower $n$ bits and store the elements based on the upper $n-1$ bits. The hashing is performed following Equation~\eqref{eq:hashing},

\begin{equation}
\label{eq:hashing}
H(x) = \frac{x}{2^n}
\end{equation}
where $n$ is the number of bits shifted.

In case of collision, we resolve the conflict by adding $1$ to the position tag, thus offsetting the storage location by $1$ towards the right. We repeat this collision resolution until an empty space is found on the hashtable (Hashtable walk). To prevent data races, we use \textit{atomic compare and exchange} instructions to test for empty locations in the hashtable. This collision resolution is shown in Figure \ref{fig:collision_resolution_upper}.

The use of the upper bits (the high-order bits) for hashing preserves the partial products' sorted order in the hashtable. Whenever collisions occur, the hashtable walk disrupts this order, making the table semi-sorted (most elements will be in sorted order, with only a few outliers).
To merge the partial products, we employ a simple \textit{atomic fetch and add} instruction to add partial products together.
\begin{figure}[htbp]
	\centering
	\includegraphics[width=1.0\textwidth]{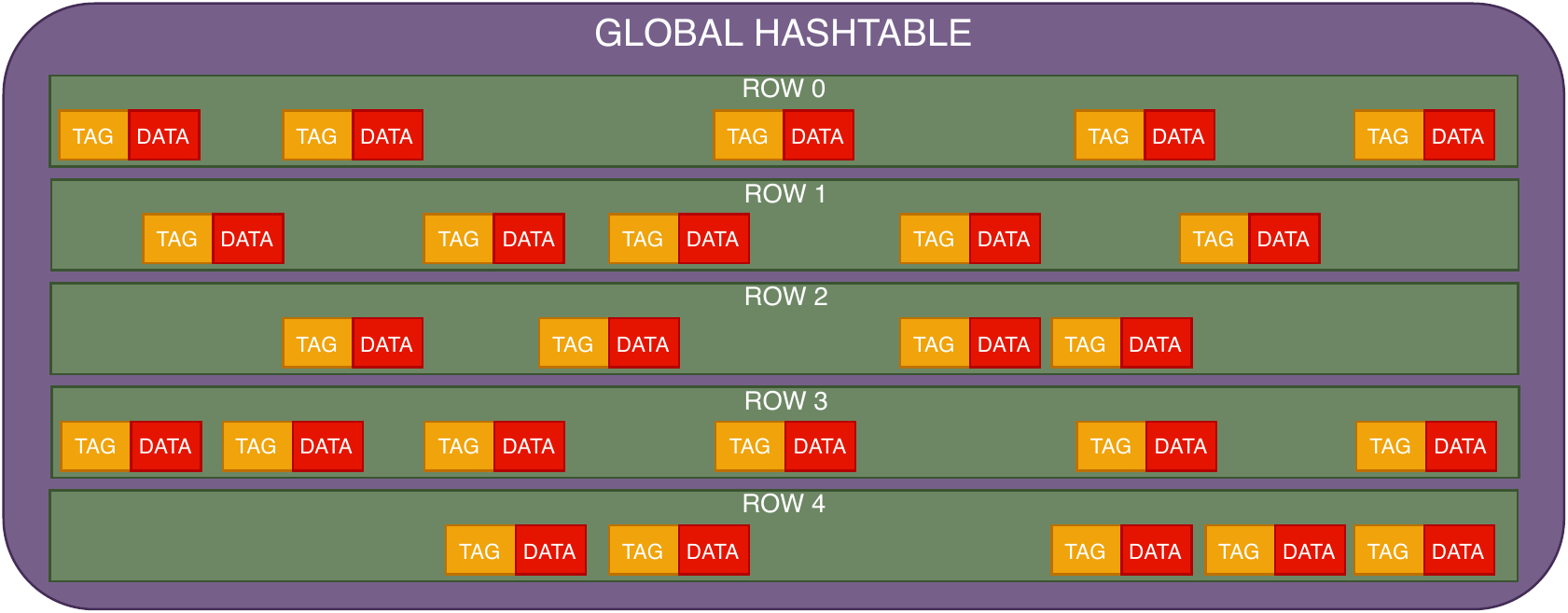}
	\caption[Tag-Data Hashtable]{\label{fig:tag_data_hashtable} Tag-Data Hashtable.}
\end{figure}
Our hashtable, with its $tag$ and $data$ pairs, is shown in Figure~\ref{fig:tag_data_hashtable}. This hashtable is stored in \ac{SPAD} memory for quick updates by the atomic instructions.

The pseudo-code for the entire hashing phase is shown in the Algorithm~\ref{alg:hashing}.
\begin{algorithm}[htbp]
\label{alg:hashing}
\linespread{0.4}\selectfont
\SetAlgoLined
\caption{SMASH HASHING}
\DontPrintSemicolon
\tcp*[h]{READ PHASE}\;
\While{Till you reach end of window} {
\tcp*[h]{Atomically distribute work to each thread}\;
$token \leftarrow$ Each thread will receive one unique token\;
\eIf{$token\_id\ \%\ 2 = 0$}{
$row\_begin \leftarrow A\_col\_ptr\_copy\_1[\frac{token\_id}{2}]$\;
}{
$row\_begin \leftarrow A\_col\_ptr\_copy\_2[\frac{token\_id}{2}]$\;
}
$row\_end \leftarrow A\_col\_ptr[\frac{token\_id}{2} + 1]$\;
\For{$i \leftarrow$ Iterate from $row\_begin$ to $row\_end$}{
\If{Check if we are within our assigned window}{
$col\_begin \leftarrow B\_row\_ptr[\frac{token\_id}{2}]$\;
$col\_end \leftarrow B\_row\_ptr[\frac{token\_id}{2} + 1]$\;
\eIf{$token\_id\ \%\ 2 = 0$}{
\tcp*[h]{Hash EVEN Section}\;
}{
\tcp*[h]{Hash ODD Section}\;
}
}
}
}	
$A\_col\_ptr\_copy\_1$ and $A\_col\_ptr\_copy\_2$ will now reflect new positions\;
\end{algorithm}

\begin{algorithm}[htbp]
\label{alg:hashing:even}
\linespread{0.4}\selectfont
\SetAlgoLined
\caption{SMASH HASHING Even Section}
\DontPrintSemicolon

\For{$k \leftarrow$ Iterate from \textcolor{red}{ $col\_begin$ } to \textcolor{red}{ $\frac{col\_end - col\_begin}{2}$ } }{
\tcp*[h]{Multiply element from $mat\_A$ with that from $mat\_B$ and store its tag and value}\;
$tag \leftarrow$ $X$ coordinate from $mat\_A$ element and $Y$ coordinate from $mat\_B$ element\;
\tcp*[h]{Hash the Tag}\;
$tag \leftarrow tag >> hash\_shift$\;
\eIf{$SPAD\_tag[tag] = EMPTY$} {
$SPAD\_tag[tag] \leftarrow tag$  \tcp*[h]{Store Tag on scratchpad}\;
$SPAD\_val[tag] \leftarrow value$  \tcp*[h]{Store Value on scratchpad}\;
}
{
\eIf{$SPAD\_tag[tag] = tag$}
{
$SPAD\_val[tag] += value$  \tcp*[h]{Accumulate Value}\;
}
{
\tcp*[h]{Probe for empty space on Scratchpad}\;
}
}
}
\end{algorithm}

\begin{algorithm}[htbp]
\label{alg:hashing:odd}
\linespread{0.4}\selectfont
\SetAlgoLined
\caption{SMASH HASHING Odd Section}
\DontPrintSemicolon

\For{$k \leftarrow$ Iterate from \textcolor{red}{ $col\_end$ } to \textcolor{red}{ $\frac{col\_end - col\_begin}{2}$ } }{
\tcp*[h]{Multiply element from $mat\_A$ with that from $mat\_B$ and store its tag and value}\;
$tag \leftarrow$ $X$ coordinate from $mat\_A$ element and $Y$ coordinate from $mat\_B$ element\;
$tag \leftarrow tag >> hash\_shift$\;
\eIf{$SPAD\_tag[tag] = EMPTY$} {
$SPAD\_tag[tag] \leftarrow tag$  \tcp*[h]{Store Tag on scratchpad}\;
$SPAD\_val[tag] \leftarrow value$  \tcp*[h]{Store Value on scratchpad}\;
}
{
\eIf{$SPAD\_tag[tag] = tag$}
{
$SPAD\_val[tag] += value$  \tcp*[h]{Accumulate Value}\;
}
{
\tcp*[h]{Probe for empty space on Scratchpad}\;
}
}
}
\end{algorithm}

\subsection{Write-back Phase}
The write-back phase moves the partial products from the hashtable to their final output matrix, stored in \ac{DRAM} in the \ac{CSR} format.
In the write-back phase, first, we employ a sorting mechanism to sort the partially sorted hashtable.
We take advantage of the partially sorted hashtags for our implementation and sort them using a variation of insertion sort.
Using insertion sort also helps us merge the remaining partial products by matching their $tags$. This enables us to stream elements from a hashtable, in ascending order, directly to the output matrix present stored in \ac{DRAM} in the \ac{CSR} format. Pseudocode for this phase is as shown in Algorithm~\ref{alg:writeback}.
\begin{algorithm}[htbp]
\label{alg:writeback}
\linespread{0.5}\selectfont
\SetAlgoLined
\caption{SMASH WRITEBACK}
\DontPrintSemicolon
\tcp*[h]{WRITE PHASE}\;
\tcp*[h]{Divide SPAD into 64 equal sections}\;
$scan\_start \leftarrow tid \times \frac{Total\ Bins\ on\ SPAD}{64}$\;
\tcp*[h]{Add an offset to $scan\_start$ to take into account overflow}\;
$scan\_start \leftarrow scan\_start - OFFSET\_THRESHOLD$\;
$scan\_end \leftarrow tid \times (\frac{Total\ Bins\ on\ SPAD}{64} + 1)$\;
$index \leftarrow 0$ \tcp*[h]{Counter to keep track of last written element on C matrix}\;
\For{$i \leftarrow$ Iterate from $scan\_start$ to $scan\_end$}{
\If{$SPAD\_tag[i] \neq EMPTY$}{
\tcp*[h]{Match value on minimum heap tree}\;
$mat\_C\_tag[tid][index] \leftarrow SPAD\_tag[i]$\tcp*[h]{Copy tag from scratchpad to DRAM}\;
$mat\_C\_val[tid][index] \leftarrow SPAD\_val[i]$\tcp*[h]{Copy value from scratchpad to DRAM}\;
}
}
\end{algorithm}

\begin{figure}[htbp]
	\centering
	\includegraphics[width=1.0\textwidth]{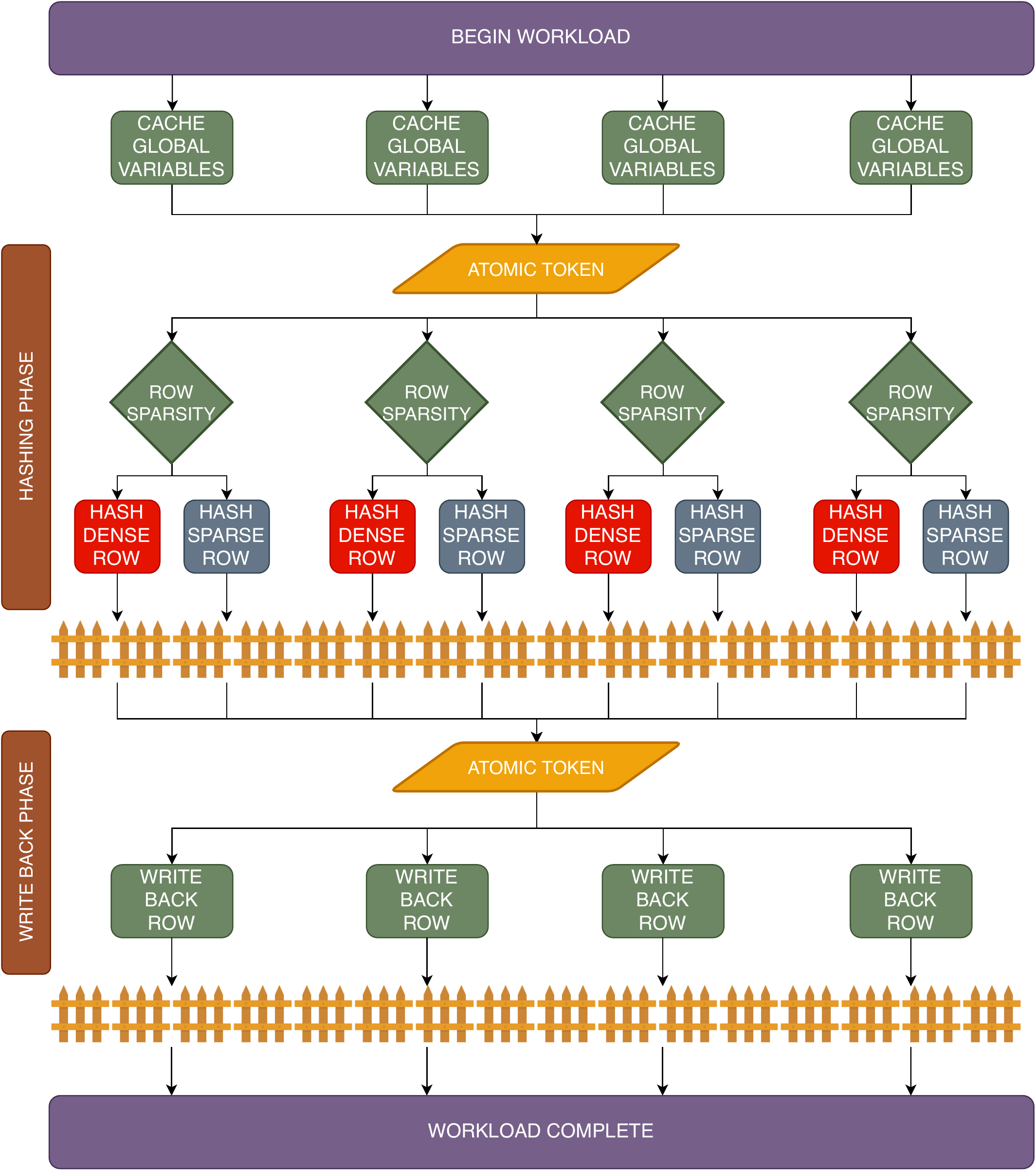}
	\caption[SMASH Algorithm]{\label{fig:smash_algorithm} SMASH Algorithm}
\end{figure}

An overview of the entire SMASH algorithm is presented in Figure \ref{fig:window_distro} and Figure \ref{fig:smash_algorithm}.

\section{SMASH Version 2: Tokenization}
\label{chap:smash:section:tokenization}
The previous version allows one row of the input matrix to be assigned to one thread on the \ac{PIUMA} block. This leads to a considerable imbalance in work across all threads in a block. The \ac{SpGEMM} datasets are notorious for encountering workload imbalance during kernel execution.

We tackle the issue of workload imbalance by adding a dynamic work scheduler layer in our Hashing phase.
Instead of statically allocating rows to threads in a round-robin fashion, we adopt the Producer-Consumer for model row allocation.
The dynamic row allocation works as follows:
\begin{enumerate}
	\item Generate two tokens for every single row present in the window.
	\item Each \ac{PIUMA} thread polls for a single token. Thus, every row is allocated 2 \ac{PIUMA} threads.
	\item These 2 \ac{PIUMA} threads start hashing the row. The first thread starts from the beginning of the row and hashes the first half of the row (i.e., the even section). The second thread does the same over the second half of the row (i.e., the odd section).
	\item Partial products from both threads are hashed into a common hashtable, stored in the \ac{SPAD} memory.
	\item When all of the tokens have been polled, the window execution is completed. 
\end{enumerate}
The split of workload between even and odd sections can be seen in  Algorithms~\ref{alg:hashing},~\ref{alg:hashing:even}, and~\ref{alg:hashing:odd}.

Despite the overhead of polling tokens, tokenization produces great speedup over static allocation, as it achieves a near-perfect distribution of workload across threads. More details of the performance are presented in the following chapter. 

\begin{figure}[htbp]
	\centering
	\includegraphics[width=0.8\textwidth]{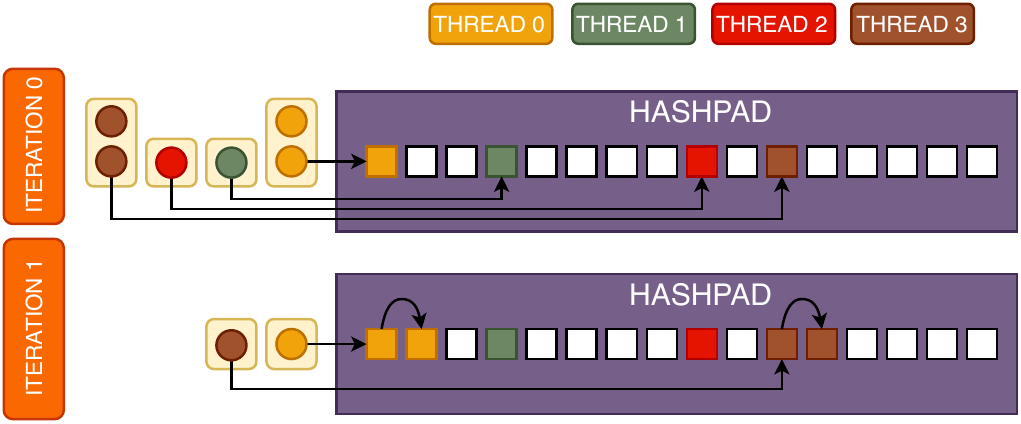}
	\caption[Hashing on low-order bits.]{\label{fig:hash_lower_bits} Hashing on low-order bits.}
\end{figure}

Another optimization carried out in \ac{SMASH} Version 2 is the change of hashing bits.
In \ac{SMASH} Version 1, we used the high-order bits for hashing in the hashtable. The downside of using high-order bits is that if two elements with their $tag$ values close to each other need to be hashed, they will end up hashing to the same position, thus following the collision routine. Hashing on high-order bits groups clusters of adjacent elements together. Instead, in this version, we chose to hash on low-order bits by setting the top $n$ high-order bits to $zero$. 
Using low-order bits evenly distributes a cluster of elements over the entire hashtable, thus sharply reducing the number of collisions. This can be seen in Figure~\ref{fig:hash_lower_bits}

The disadvantage of using low-order bits is that the order of hashing is no longer preserved. The hashtable is no longer partially sorted, as in the case of the previous version.
We overcome this problem by merging all partial products of the same $tag$ before writing them to the hashtable.
Even though the order is not preserved and the output matrix in \ac{CSR} format is not sorted, the correctness of the solution is maintained, as all partial products are properly merged.

\section{SMASH Version 3: Fragmenting Memory}
Previous sections describe how \textit{Atomic hashing} removes redundant accesses to the partial product matrices and \textit{Tokenization} balances workload across \ac{PIUMA} threads.

This section describes the integration of the \ac{DMA} engine in the \ac{SMASH} algorithm by fragmenting the \ac{SPAD} memory.
The \ac{DMA} engine provides the \ac{PIUMA} system the capability to move data independently within its global address space while the \ac{PIUMA} blocks work on other segments of the algorithm.

\begin{figure}[htbp]
	\centering
	\includegraphics[width=0.9\textwidth]{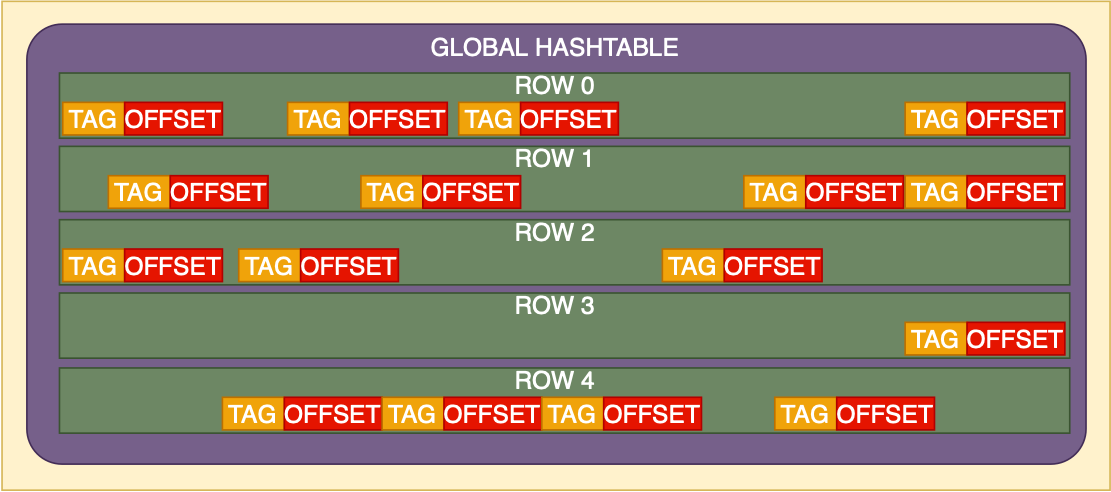}
	\caption[Tag-Offset Hashtable]{\label{fig:v3_dram_layout} Tag-Offset Hashtable in DRAM.}
\end{figure}

We incorporate a \textit{copy} instruction to move data from the \ac{SPAD} to \ac{DRAM}, as well as a \textit{scatter} instruction to prepare the \ac{DRAM} for the next window.
To use the \ac{DMA} engine efficiently, we make modifications to our previous version of \ac{SMASH}. These modifications included:
\begin{enumerate}
	\item In addition to hashing on a common global hashtable, the \ac{PIUMA} threads also maintain a private local array that stores the $tag$ values in a dense array.
	\item Instead of storing the hashtable in the \ac{SPAD}, we store the hashtable in \ac{DRAM}, a memory that has lower bandwidth but more available space.
	\item As every element gets hashed on the global hashtable, we store its position in the hashtable in a separate array in \ac{SPAD}, called the $offset\_array$.
\end{enumerate}
\begin{figure}[htbp]
	\centering
	\includegraphics[width=0.7\textwidth]{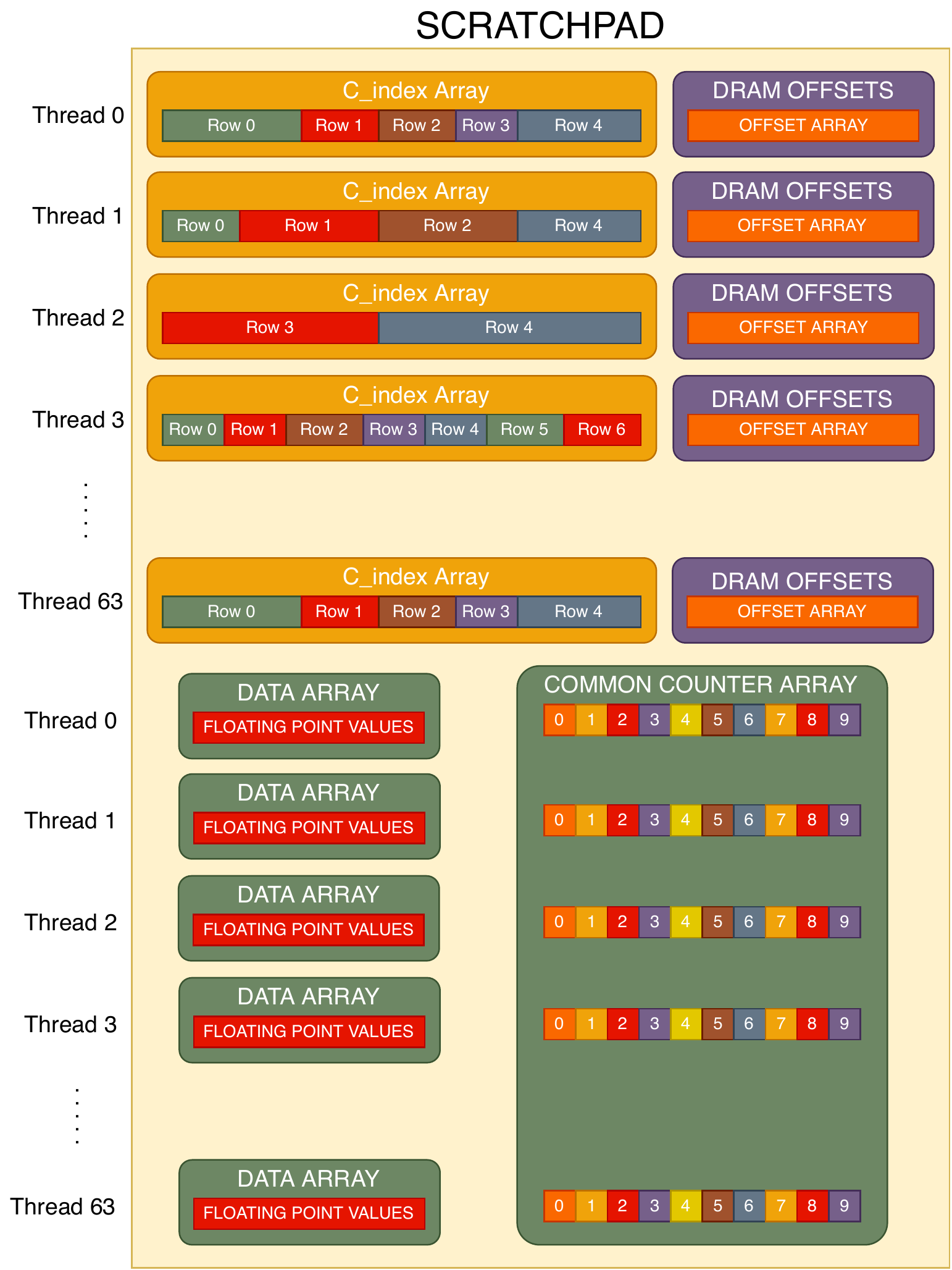}
	\caption[Window Distribution Algorithm]{\label{fig:v3_spad_layout} SPAD array layout}
\end{figure}
The layout of the \ac{SPAD} in \ac{SMASH} V3 can be best described by Figure~,\ref{fig:v3_spad_layout} and the layout of \ac{DRAM} can be best described by Figure~\ref{fig:v3_dram_layout}.

\ac{SMASH} Version 1 and \ac{SMASH} Version 2 store $tag$ and $data$ values in a hashtable, scattered across a large array.
\ac{SMASH} Version 3, on the other hand, stores the $tag$ values and $data$ values in dense arrays. These dense arrays can then be easily transferred to the \ac{DRAM} by simple $copy$ instructions. Also, since the \ac{DMA} engine runs in parallel with the \ac{MTC}s, the \ac{PIUMA} threads will not have to spend cycles moving these dense arrays from the \ac{SPAD} to \ac{DRAM}.

In the next chapter, we will look deeper into the performance results of each \ac{SMASH} implementation and the advantages gained by optimizing each version.

%% file: tex/results.tex

\chapter{Results}
\label{chap:results}

This chapter begins by providing more details on the evaluation methodology used in our experiments. Then, following this methodology, we compare the performance of different versions of our \ac{SMASH} algorithm, thus providing insights on the improvements offered by, and the challenges remaining in, each implementation.

\section{Experimental Methodology}

For our experiments, we chose the R-MAT synthetic sparse matrix dataset~\cite{chakrabarti2004r, hubschle2020linear}. In prior work~\cite{chakrabarti2004r}, Chakrabarti et al. proposes a synthetic graph generator that closely represents real-world graphs in multiple disciplines. In addition to their graph generator being fast, multithreaded, and robust, they successfully simulate the famous Erdos-Renyi model~\cite{daudin2008mixture}, providing a measure of the uniform probability distribution of each possible independent edge of a graph. The RMAT sparse matrices exhibit irregular sparsity patterns with a power-law distribution of non-zeros, making them notoriously difficult to balance between threads. A synthetic data generator also allows us to create graphs of specific required dimensions and varying sparsity patterns for analysis.

We generate two $16K \times 16K$ matrices using the R-MAT generator and multiply them with each other using the row-wise multiplication method. We evaluate the performance of all 3 of our kernel implementations on the same input matrices using our simulator and report various performance metrics in the next section.

\section{Dataset Arithmetic Intensity}
Before we dive deeper into the analytics of our \ac{SpGEMM} implementation, it is worthwhile to explore the arithmetic intensity of multiplying two sparse matrices. The characteristics of the matrices used in this thesis are shown in Table~\ref{table:data_metrics}.

\begin{table}[ht]
	\centering 
	\begin{tabular}{l c c c} 
		\hline\hline 
		\textbf{Matrix} & \textbf{Dimensions} & \textbf{Total Non-zeros} & \textbf{Sparsity} \\ [0.5ex] 
		\hline 
		Input Matrix A & 16,384 * 16,384 & 254,211 & 99.9\% \\
		Input Matrix B & 16,384 * 16,384 & 254,211 & 99.9\% \\
		Output Matrix C & 16,384 * 16,384 & 5,174,841 & 98.1\% \\ [1ex] 
		\hline 
	\end{tabular}
	\caption{Input and output data characteristics used in this thesis.} 
	\label{table:data_metrics} 
\end{table}

The arithmetic Intensity of \ac{SpGEMM} is computed as the ratio of the total number of floating-point operations to the number of total data movement operations (reported in bytes)~\cite{gu2020bandwidth}. An AI value of $0.09$ or $\frac{9}{100}$ means for 9 floating-point operations, at least 100 bytes of data need to be moved.The arithmetic intensity (AI) for multiplying sparse matrix A with sparse matrix B to produce output matrix C is given by equation~\ref{eq:ai_spgemm}:

\begin{equation}
\label{eq:ai_spgemm}
AI \leq \frac{nnz(C)*cf}{[nnz(A)+nnz(B)+nnz(C)]*b} \leq \frac{cf}{b}
\end{equation}

where $nnz$ is the total number of non-zeros in the matrix, $b$ is the total number of bytes required to store one element of the input matrix, $cf$ is the compression factor computed as a ratio of FLOPs to nonzeros in the output matrix as seen in Equation~\ref{eq:ai_cf}. 

\begin{equation}
\label{eq:ai_cf}
cf = \frac{flop}{nnz(C)}
\end{equation}

\begin{table}[ht]
	\centering 
	\begin{tabular}{l c c c c} 
		\hline\hline 
		\textbf{Matrix Parameters} & \textbf{Data Type} & \textbf{Elements} & \textbf{Size (Bytes)} & \textbf{Size (in KB)} \\ [0.5ex] 
		\hline 
		Row Pointer & INT 4 Bytes & 16,385 & 65,540 & 64 KB \\
		Column Index & INT 4 Bytes & 2,54,211 & 1,016,844 & 993 KB \\
		Data Array & Double 8 Bytes & 2,54,211 & 2,033,688 & 1,986 KB \\
		\hline 
		Total & - & 5,683,263 & 3,116,072 & 3,043 KB \\ [1ex]
		
		\hline 
	\end{tabular}
	\caption{CSR matrix arrays for input matrices A and B.} 
	\label{table:data_sizes_input} 
\end{table}

\begin{table}[ht]
	\centering 
	\begin{tabular}{l c c c c} 
		\hline\hline 
		\textbf{Matrix Parameters} & \textbf{Data Type} & \textbf{Elements} & \textbf{Size (Bytes)} & \textbf{Size (in KB)} \\ [0.5ex] 
		\hline 
		Row Pointer & INT 4 Bytes & 16,385 & 65,540 B & 64 KB \\
		Column Index & INT 4 Bytes & 5,174,841 & 20,699,364 & 20,214 KB \\
		Data Array & Double 8 Bytes & 5,174,841 & 41,398,728 & 40,428 KB \\
		\hline 
		Total & - & 10,366,067 & 62,163,632 & 60,706 KB \\ [1ex]
		
		\hline 
	\end{tabular}
	\caption{CSR matrix arrays for the output matrix C.} 
	\label{table:data_sizes_output} 
\end{table}
\FloatBarrier

In our case, we consider one particular example to compute $cf$ and $AI$ using data metrics, as shown in Table~\ref{table:data_metrics}, Table,\ref{table:data_sizes_input} and Table~\ref{table:data_sizes_output}.

For our implementation, we compute the compression factor, $cf = 1.23$. We further compute our arithmetic intensity $AI$ using this $cf$.
The arithmetic intensity of our \ac{SMASH} Version 3 implementation is $AI = 0.09$.

\section{DRAM Performance}

Both the input matrices and the output matrix are stored in \ac{DRAM}. The \ac{DRAM} bandwidth is a measure of the rate at which the input matrices are read, and the output matrices are written~\cite{4336216}.
The \ac{DRAM} bandwidth is considered a bottleneck for \ac{SpGEMM} implementations~\cite{tithi2021performance}. We compare the \ac{DRAM} bandwidth utilization for all our \ac{SMASH} implementations. DRAM bandwidth helps us decide if our \ac{SpGEMM} kernel is memory-bound or compute-bound. Changes in bandwidth utilization over time also help us narrow down algorithm phases that produce a bottleneck.
Considering we are using a row-wise product approach, the \ac{DRAM} is utilized to read the row pointers, column indices, and data values for input matrices $A$ and $B$.
As version 3 of our \ac{SpGEMM} implementation stores the hashtable in \ac{DRAM}, this hashtable contributes to the \ac{DRAM} bandwidth demands as well.
The \ac{DRAM} bandwidth demands are compared in Table~\ref{table:dram_compare}. 

\begin{table}[ht]
	\centering 
	\begin{tabular}{l c} 
		\hline\hline 
		\textbf{SMASH Versions} & \textbf{DRAM Bandwidth} \\ [0.5ex] 
		\hline 
		Version 1 & 55.2\% (3.03 GB/s) \\
		Version 2 & 73.9 \% (4.06 GB/s)\\
		Version 3 & 95.9\% (5.26 GB/s) \\ [1ex]
		\hline 
	\end{tabular}
	\caption{Aggregated DRAM bandwidth demands.} 
	\label{table:dram_compare} 
\end{table}

\section{Cache Performance}

Cache performance and utilization play a crucial role in the speed achieved by our \ac{SpGEMM} kernels. We maintain temporal locality across the first input matrix elements and spatial locality across the second input matrix.
This reuse of elements from the first matrix, combined with the access to neighboring elements from the second matrix, allows us to achieve high data-cache hit rates.
The L1 data-cache hit rates for all 3 versions of our \ac{SMASH} algorithm are presented in Table~\ref{table:cache_compare}.

\begin{table}[ht]
	\centering 
	\begin{tabular}{l c} 
		\hline\hline 
		\textbf{SMASH Versions} & \textbf{L1 Data Cache Hit Rate} \\ [0.5ex] 
		\hline 
		Version 1 & 88.7\% \\
		Version 2 & 92.2\% \\
		Version 3 & 94.1\% \\ [1ex]
		\hline 
	\end{tabular}
	\caption{Cache performance of our 3 SMASH implementatinos.} 
	\label{table:cache_compare} 
\end{table}

\section{Workload Distribution}
We achieved a significant performance improvement when leveraging \textit{tokenization} in \ac{SMASH} Version 2. This leads to a near-perfect balance of workload across \ac{PIUMA} threads. This avoided idle \ac{PIUMA} cores, with individual cores waiting for other cores to finish. This, in turn, significantly boosted the \ac{IPC}, as shown in the following section.

\begin{figure}[htbp]
	\centering
	\includegraphics[width=0.8\textwidth]{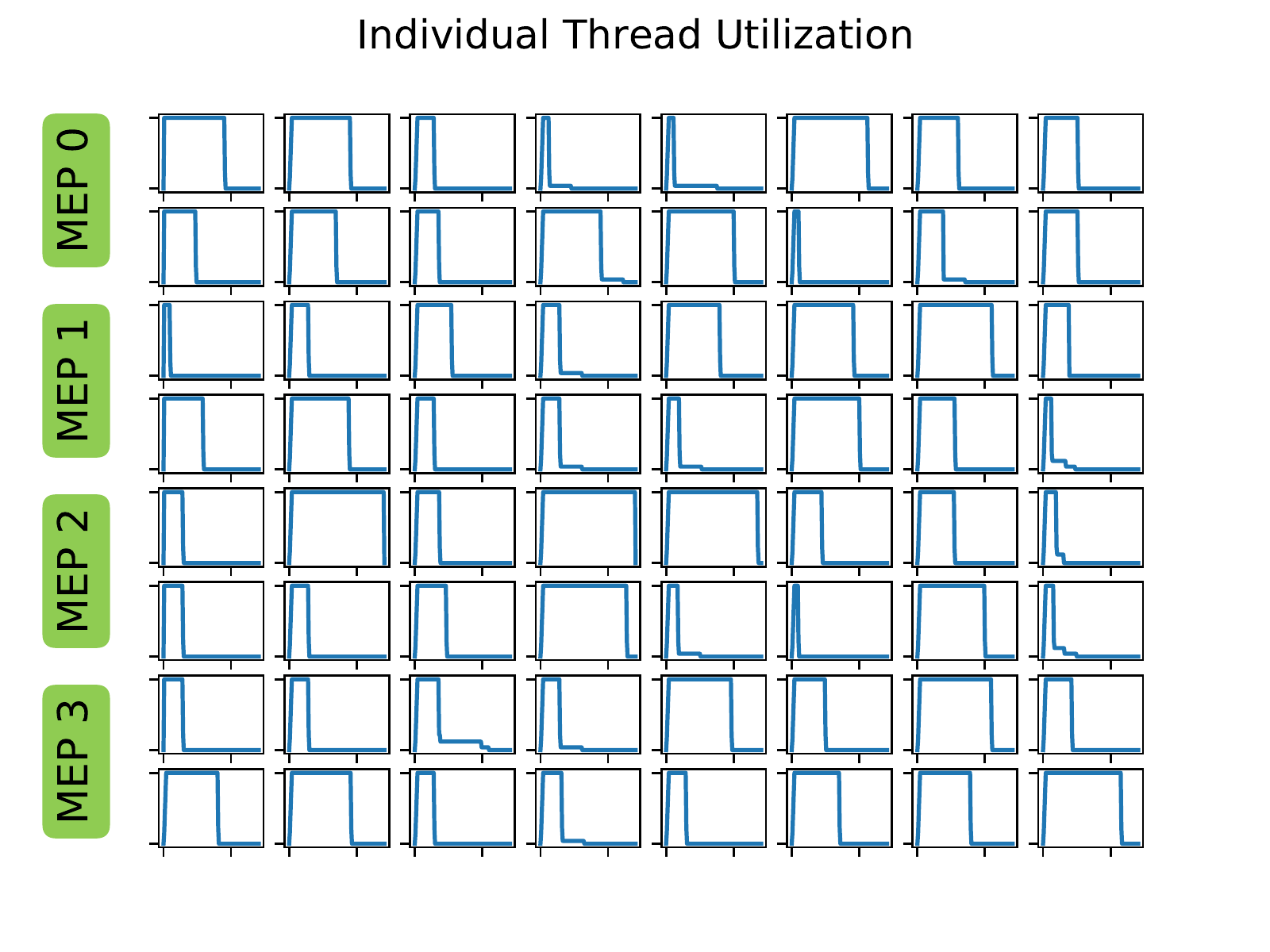}
	\caption[SMASH V1: Thread utilization plots for unbalanced workload.]{\label{fig:thread_utilization_unbalanced_workload} SMASH V1: Thread utilization plots for unbalanced workload.}
\end{figure}

\begin{figure}[htbp]
	\centering
	\includegraphics[width=0.8\textwidth]{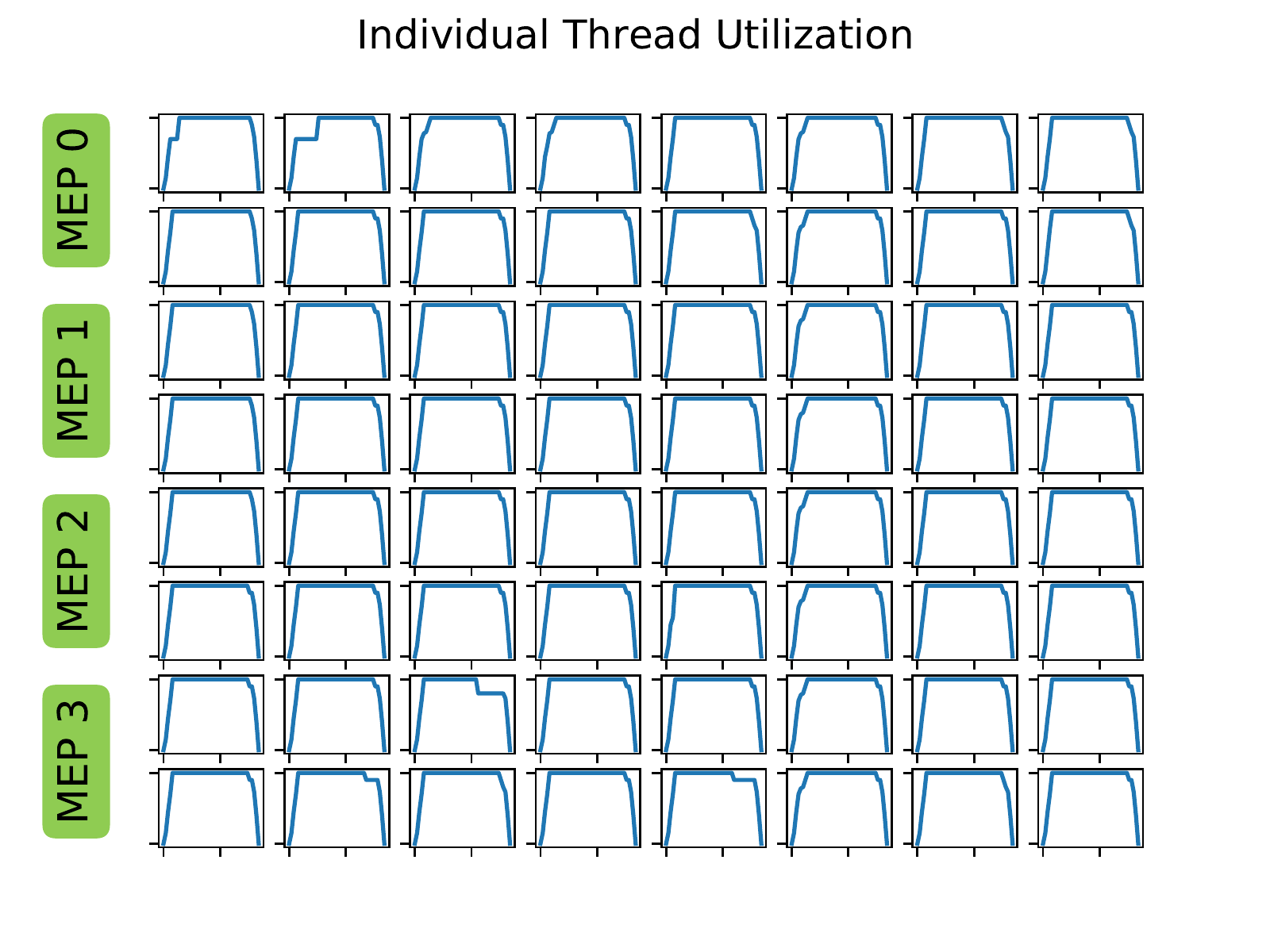}
	\caption[SMASH V2: Thread utilization plots for balanced workload.]{\label{fig:thread_utilization_balanced_workload} SMASH V2: Thread utilization plots for balanced workload.}
\end{figure}

\begin{figure}[htbp]
	\centering
	\includegraphics[width=0.9\textwidth]{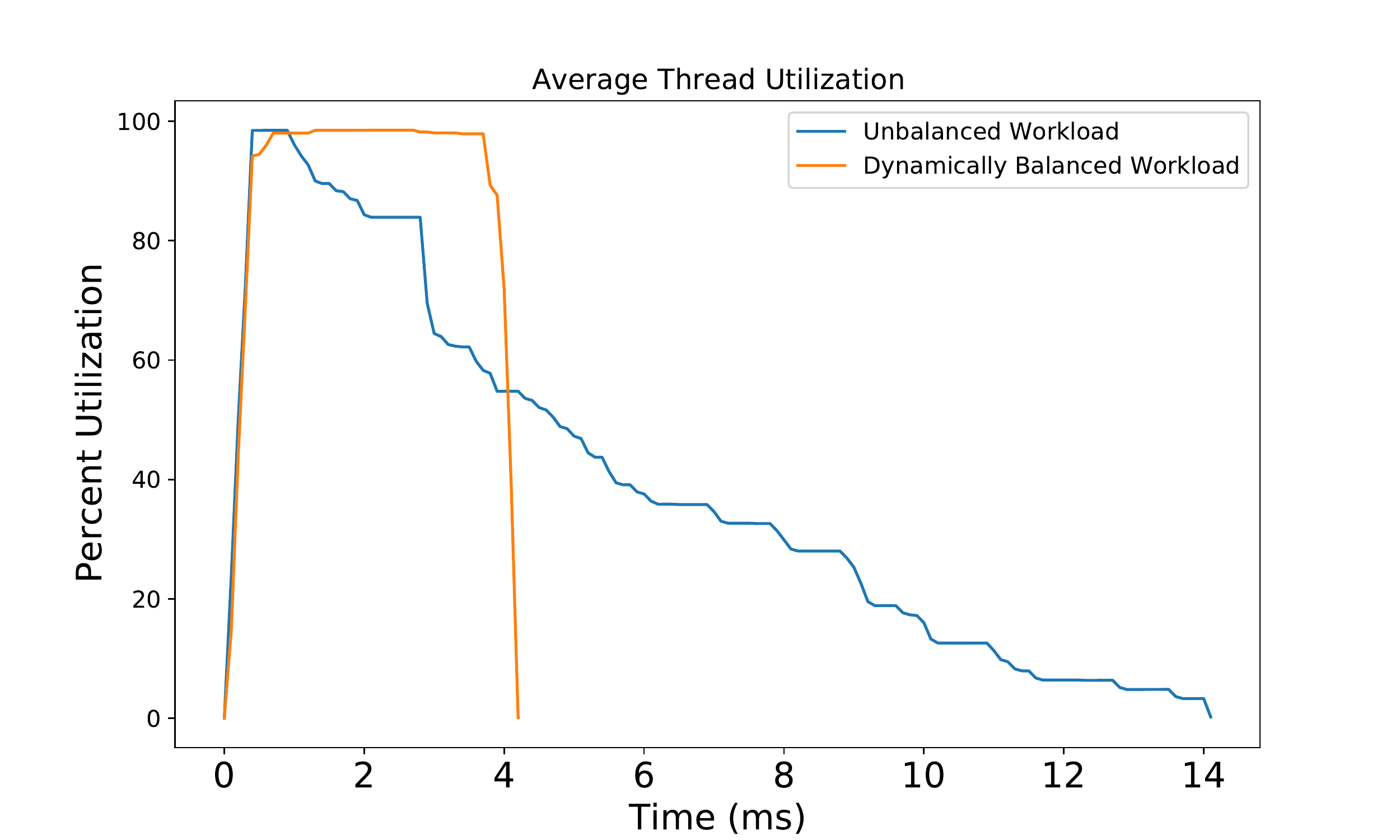}
	\caption[Average thread utilization.]{\label{fig:avg_thread_utilization} Average thread utilization.}
\end{figure}

\begin{figure}[htbp]
	\centering
	\includegraphics[width=0.9\textwidth]{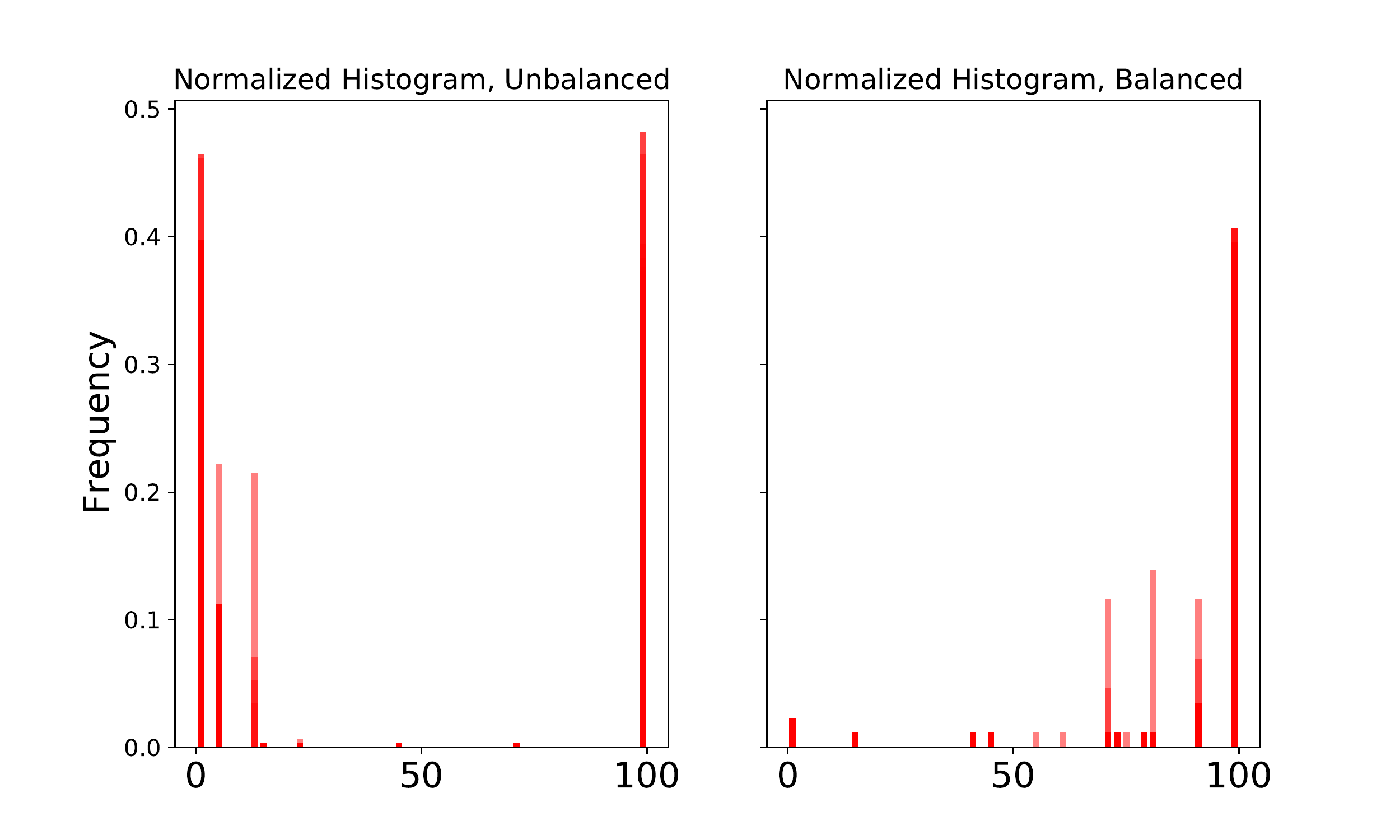}
	\caption[Thread utilization histogram comparison between balanced and unbalanced workloads.]{\label{fig:normalized_utilization_histogram} Thread utilization histogram comparison between balanced and unbalanced workloads.}
\end{figure}

We analyze the performance of \ac{SMASH} Version 1 and \ac{SMASH} Version 2 on a single window, using a single \ac{PIUMA} block.  Results are shown in Figures~\ref{fig:thread_utilization_unbalanced_workload}, \ref{fig:thread_utilization_balanced_workload}, \ref{fig:avg_thread_utilization} and \ref{fig:normalized_utilization_histogram}.

Figures~\ref{fig:thread_utilization_unbalanced_workload} and~\ref{fig:thread_utilization_balanced_workload} provide information on the utilization of each thread in a block, measured over time. 
These figures represent the thread utilization, with the x-axis plotting the time in milliseconds and the y-axis reporting the associated thread utilization.

Figure~\ref{fig:thread_utilization_unbalanced_workload} shows thread utilization of SMASH V1 kernel. As observed in this figure, some of the threads do not achieve high thread utilization, indicating that the multi-threaded core is underutilized as some of the threads stall during execution (they stall on barriers, waiting for other threads to complete).

Figure~\ref{fig:thread_utilization_balanced_workload} shows the same workload on the SMASH V2 kernel. All threads in this figure achieve close to $100\%$ thread utilization. This shows that our later implementation has mitigated the cause of under-utilization of the multi-threaded cores.

Figure~\ref{fig:avg_thread_utilization} reports the average thread utilization for each workload. Using dynamic allocation prevents threads from stalling while waiting for other threads to complete, thus maintaining a higher \ac{IPC} value.

In Figure~\ref{fig:normalized_utilization_histogram} we provide a normalized histogram to report on thread utilization, showing the performance improvement in \ac{SMASH} V2 as compared to \ac{SMASH} V1. The balanced workload shown on the right exhibits more threads achieve nearly 100\% utilization, as opposed to the unbalanced workload, where multiple threads idling.

With the use of $tokenization$, we managed to not only distribute workload evenly across all threads, but we also end up reducing the overall execution time required for this window from 14.15 ms to 4.09 ms, despite the presence of overhead introduced when creating and polling tokens.

\section{Instruction Throughput}

\ac{IPC} describes the total number of instructions being executed in the system for every cycle.
We compare our \ac{SMASH} implementations by comparing aggregate \ac{IPC} over the entire execution of the workload while considering all \ac{PIUMA} threads.
Ideally, the max value of the \ac{IPC} that can be achieved is equal to the number of \ac{MTC}s present in the block.

\begin{equation}
Aggregate\ IPC = \frac{Total\ Instructions\ Executed}{Total\ Cycles}
\label{eq:agg_ipc}
\end{equation}

Aggregate \ac{IPC} can be computed using the equation~\ref{eq:agg_ipc}.

\begin{table}[ht]
	\centering 
	\begin{tabular}{l c} 
		\hline\hline 
		\textbf{SMASH Versions} & \textbf{Aggregate IPC} \\ [0.5ex] 
		\hline 
		Version 1 & 0.9 IPC \\
		Version 2 & 1.7 IPC \\
		Version 3 & 2.3 IPC \\ [1ex]
		\hline 
	\end{tabular}
	\caption{Aggregate IPC Comparisons} 
	\label{table:ipc_compare} 
\end{table}

We provide the aggregate \ac{IPC} values for each of our 3 \ac{SMASH} implementations in Table~\ref{table:ipc_compare}.

\section{Application Speedup}

We simulate the total time required by each of our \ac{SMASH} versions on our interval simulator, simulating the \ac{PIUMA} hardware.
We consider the time required to run the entire \ac{SpGEMM} workload on a single \ac{PIUMA} block.
The runtime comparison for all 3 versions of \ac{SMASH} are presented in Table~\ref{table:runtime_compare}.

\begin{table}[ht]
	\centering 
	\begin{tabular}{l c c} 
		\hline\hline
		\textbf{SMASH Versions} & \textbf{Runtimes} & \textbf{Speedup over Version 1} \\ [0.5ex]
		\hline
		Version 1 & 986.7 ms & 1.0$\times$ \\
		Version 2 & 432.5 ms &  2.3$\times$ \\
		Version 3 & 105.4 ms & 9.4$\times$ \\ [1ex]
		\hline
	\end{tabular}
	\caption{Runtime for an entire SpGEMM workload on 64 PIUMA threads.} 
	\label{table:runtime_compare} 
\end{table}

\section{Summarry of Results}

The SMASH V3 kernel is a state-of-the-art SpGEMM implementation built on the PIUMA architecture. It employs various optimizations over previous SpGEMM kernel implementations, as well as previous SMASH versions. The new kernel is capable of utilizing $95.9\%$ of DRAM bandwidth, nearly saturating the available bandwidth. With the deployment of the producer-consumer model, the SMASH V3 kernel is able to deliver almost $100\%$ multi-threaded core utilization. This optimization translates to a $9.4\times$ speedup, as well as a $155\%$ increase in instruction throughput, as seen in Tables~\ref{table:ipc_compare} and~\ref{table:runtime_compare}.

%% file: tex/conclusion.tex

\chapter{Conclusions and Future Work}
\label{chap:conclusions}


\ac{SpGEMM} workloads are well known to test the limits of both hardware and software. The software kernel implementation can play a key role in the resulting irregular memory access pattern.
For this reason, most general-purpose architectures, including CPUs and GPUs, typically fail to achieve high speedups when executing \ac{SpGEMM}-based applications.

In this work, we described the many uses of \ac{SpGEMM} kernels, helping to motivate the need for domain-specific architectures for such workloads.  We identified key issues that need to be addressed when designing \ac{SpGEMM} kernels.
We further investigated prior research that has pursued these same issues.  This helped shape the design of our approach pursued in this thesis while optimizing the mapping of the \ac{SpGEMM} kernel to the underlying \ac{PIUMA} architecture.  We utilized a row-wise product approach in each of our four implementations.

We explored some of the novel features present in the \ac{PIUMA} architecture, designed to tackle sparse graph and sparse matrix applications. We designed 3 different \ac{SpGEMM} kernel implementations called \ac{SMASH} for the \ac{PIUMA} system, focusing on the key features available on this accelerator, including \ac{DGAS}, networked instructions, \ac{DMA} Engines, and multi-threaded cores.

Our set of optimizations focused primarily on improving \ac{DRAM} bandwidth utilization of the \ac{SpGEMM} kernel. But an increase in \ac{DRAM} bandwidth utilization by itself is not an indication of improved performance, as multiple factors can impact the resulting performance.  Avoiding redundant reads to memory, poor reuse of input matrices, and increases in metadata size can all offset bandwidth utilization improvements.  Our \ac{SMASH} V3 kernel implementation stores the hashtable in memory instead of using an on-chip Scratchpad.  Thus, in addition to reading input matrices, the kernel also has to read intermediate partial-products from the hashtable stored on DRAM. Thus the \ac{DRAM} bandwidth is shared between the input data reads and the partial-product reads. But in addition to an increase in \ac{DRAM} bandwidth utilization, we also observed a significant speedup of Version 3 over previous versions.  We were able to achieve a speedup of $9.4\times$ over Version 1 by iterative improvements, performing tokenization, and memory defragmentation.

To summarize, the \ac{SMASH} kernel improvements are as follows.
\begin{itemize}
	\item We successfully built an implementation using atomic hashing that eliminated the need for partial product matrices in row-wise product methods, thus preventing redundant accesses to \ac{DRAM} memory.
	\item We improved workload balance by adding a layer of dynamic work allocation, leveraging a producer-consumer model.
	\item Finally, we leveraged \ac{PIUMA}'s \ac{DMA} engine, which enabled us to move data from the \ac{SPAD} to \ac{DRAM} without wasting precious cycles of the \ac{MTC}s.
\end{itemize}

\section{Contributions of this Thesis}

The main contributions of this thesis include:
\begin{itemize}
    \item An in-depth analysis of the inherent problems exhibited by sparse matrix multiplication kernels.
    \item A comparison study of prior architectures that support \ac{SpGEMM} workloads.
    \item A comparative study on previous implementations of \ac{SpGEMM} kernels.
    \item An architectural overview of Intel's novel PIUMA graph accelerator.
	\item A state-of-the-art \ac{SpGEMM} kernel implementation that uses the features present in the PIUMA accelerator architecture to speedup sparse matrix operations.
\end{itemize}

\section{Future Work}
Sparse matrix-matrix multiplication kernel optimizations will continue to be an active research area. A key problem to deal with in any SpGEMM kernel implementation is the resulting workload imbalance. In our implementation, we explored applying a uniform work distribution by estimating floating-point operations based on the number of non-zeros in each row. Although this method improves the algorithm's performance for our dataset, it leaves room for optimization for other sparsity patterns.

In our work, to store and merge partial products, we employed an in-memory hashtable. Such a hashtable allows us to ensure that the partial products are merged immediately as they are produced. One of the drawbacks of using hashtables is that they can cause memory hotspots. Based on the hashing mechanism in our implementation, we used either the high-order bits or low-bits bits for hashing. This resulted in some sparsity patterns to generate hotspots (multiple elements getting hashed to the same hash class) in our hashtable. Such patterns will cause our algorithm to run the collision resolution subroutine, leading to degraded performance. In our next iteration, we plan to avoid collisions by incorporating a better hashing algorithm, one that is not solely based on restricting the bits selected.  We will consider a dynamic hashing algorithm, developing one that can  adapt to different sparsity patterns.

The \ac{PIUMA} architecture provides a rich platform where we can further explore the acceleration of linear algebra operations.  We want to extend this work well beyond the present focus on the \ac{SpGEMM} kernel. We intend to explore other linear algebra subroutines (GraphBLAS) and consider how to optimize performance given the unique features of this architecture.